\documentclass[reqno]{amsart}
\usepackage{amssymb}
\usepackage{upref}

\newcommand{\bbN}{{\mathbb{N}}}

\newcommand{\bbC}{{\mathbb{C}}}
\newcommand{\bbCinf}{{\mathbb{C}_{\infty}}}
\newcommand{\Pinfmin}{{P_{\infty_{-}}}}
\newcommand{\Pinfplus}{{P_{\infty_{+}}}}

\newcommand{\calF}{{\mathcal F}}
\newcommand{\calK}{{\mathcal K}}

\newcommand{\no}{\nonumber}
\newcommand{\lb}{\label}
\newcommand{\bi}{\bibitem}
\newcommand{\f}{\frac}
\newcommand{\dott}{\,\cdot\,}
\newcommand{\Oh}{O}

\allowdisplaybreaks
\numberwithin{equation}{section}

\newtheorem{theorem}{Theorem}[section]
\newtheorem{lemma}[theorem]{Lemma}

\theoremstyle{definition}
 
\newtheorem{example}[theorem]{Example}

\newtheorem{tabl}[theorem]{Table}

\theoremstyle{remark}
\newtheorem{remark}[theorem]{Remark}

\newcommand{\abs}[1]{\lvert#1\rvert}

\begin{document}
\title[Darboux-type transformations]{Darboux-type 
transformations \\
and hyperelliptic curves} 

\author[Gesztesy]{Fritz Gesztesy}
\address{Department of Mathematics,
University of Missouri,
Columbia, MO 65211, USA}
\email{fritz@math.missouri.edu}
\urladdr{http://www.math.missouri.edu/people/fgesztesy.html}
\author[Holden]{Helge Holden}
\address{Department of Mathematical Sciences, Norwegian 
University 
of 
Science and Technology, N--7034 Trondheim, Norway}
\email{holden@math.ntnu.no}
\urladdr{http://www.math.ntnu.no/\~{}holden/}

\dedicatory{}
\thanks{Supported in part by the Research Council of Norway 
under grant
107510/410  and the
University of Missouri Research Board grant RB-97-086.}

\begin{abstract}
We systematically study Darboux-type transformations for the KdV 
and AKNS hierarchies and provide a complete account of their 
effects 
on hyperelliptic curves associated with algebro-geometric solutions 
of these hierarchies.
\end{abstract}

\maketitle

\section{Introduction} \lb{intro}

Since the early days of completely integrable evolution equations in
the late sixties, Darboux-type transformations (cf., e.g.,
\cite{MS91} and the references therein) played an important role 
and turned out to be an integral part of (auto-)B\"acklund
transformations between soliton equations. The canonical example 
in this context is Miura's transformation
\cite{Mi68} between the Korteweg-de Vries (KdV) hierarchy and modified
Korteweg-de Vries (mKdV) hierarchy. As described in
\eqref{2.35}--\eqref{2.42}, Miura's transformation, disregarding
the time variable (i.e., considering the stationary case), is effected
by the classical Crum-Darboux transformation, which in turn is based on
factorizing the second-order Lax (one-dimensional Schr\"odinger)
operator
\begin{equation}
L=-\f{d^2}{dx^2}+V \lb{1.1}
\end{equation}
for the KdV hierarchy into a product of first-order differential
expressions plus a shift,
\begin{equation}
L=AA^+ + z_0, \quad A=\f{d}{dx}+\phi, \quad A^+=-\f{d}{dx} +\phi, 
\quad V=\phi^2+\phi_x+z_0. \lb{1.2}
\end{equation}
Assuming $V$ to satisfy one of the stationary KdV equations and
reversing the order of the two factors $A$ and $A^+$ produces a new
Lax operator $\widetilde L,$
\begin{equation}
\widetilde L=A^+A + z_0=-\f{d^2}{dx^2} + \widetilde V, \quad 
\widetilde V =\phi^2 -\phi_x +z_0, \lb{1.3}
\end{equation} 
whose potential $\widetilde V$ is a new solution of one of the
equations in the stationary KdV hierarchy. In short, the transformation
\begin{equation}
V\mapsto \widetilde V \lb{1.4}
\end{equation}
represents a Darboux transformation, or equivalently, an
auto-B\"acklund transformation of the stationary KdV hierarchy.
Incidentally, $\pm\phi$ in \eqref{1.2}, \eqref{1.3} represent
solutions of one of the stationary equations of the mKdV hierarchy and
hence 
\begin{equation}
V\mapsto\phi\mapsto-\phi\mapsto\widetilde V \lb{1.5}
\end{equation}
represents a B\"acklund transformation from the (stationary) KdV to
the mKdV hierarchy ($V\mapsto\phi$) as well as auto-B\"acklund
transformations for the KdV ($V\mapsto\widetilde V$) and mKdV hierarchy
($\phi\mapsto -\phi$). However, while $\phi$ and $-\phi$ satisfy the
identical equation(s) within the stationary mKdV hierarchy, and hence 
\begin{equation}
D=\begin{pmatrix}0 & A^+ \\ A & 0  \end{pmatrix} \mapsto 
\widetilde D =\begin{pmatrix}0 & -A \\ -A^+ & 0  \end{pmatrix} 
\lb{1.6}
\end{equation} 
represents an isospectral deformation of $D$, $V$ and $\widetilde V$ in
general do not satisfy the same stationary equation of the KdV
hierarchy, that is, 
\begin{equation}
L \mapsto \widetilde L, \lb{1.7}
\end{equation}
in general, is not an isospectral deformation of $L.$ More precisely,
each solution $V$ of (one of) the $n$th stationary KdV equations is
associated with a (possibly singular) hyperelliptic curve $\calK_n$ of
the type
\begin{equation}
\calK_n: y^2=\prod_{m=0}^{2n} (z-E_m), \quad \{E_m\}_{m=0,\dots,2n} 
\subset \bbC. \lb{1.8}
\end{equation}
Similarly, $\widetilde V$ corresponds to a curve $\widetilde
\calK_{\tilde n}$ of the type
\begin{equation}
\widetilde \calK_{\tilde n}: y^2=\prod_{m=0}^{2\tilde n} 
(z-\widetilde
E_m), \quad \{\widetilde E_m\}_{m=0,\dots,2\tilde n} \subset \bbC
\lb{1.9} 
\end{equation} 
and hence $V$ and $\widetilde V$ (resp. $L$ and $\widetilde L$) are
isospectral if and only if $\calK_n=\widetilde \calK_{\tilde n}$
(i.e., $\{E_m\}_{m=0,\dots,2n}=\{\widetilde E_m\}_{m=0,\dots,2\tilde
n},$ $n=\tilde n$). 

The principal aim of the paper is to re-examine the relation between
$\calK_n$ and $\widetilde \calK_{\tilde n}$, depending on various
choices of the  function $\phi$ in \eqref{1.2}, and to provide a
complete, yet elementary solution of this problem. Historically, the 
first attempts to link $\calK_n$ and $\widetilde \calK_{\tilde n}$ 
were made by Drach \cite{Dr18},
\cite{Dr19}, \cite{Dr19a}. The first solution
of this problem was obtained by Ehlers and Kn\"orrer in 1982 on the
basis of  purely algebro-geometric techniques. An elementary but quite
elaborate approach was recently developed by Ohmiya
\cite{ohm97} (based on two additional papers \cite{ohm95},
\cite{om93}). Our proof of Theorem~\ref{theorem2.3} appears to be the
only elementary and short one at this point. 

It seems worthwhile pointing out that the curves \eqref{1.8} and
\eqref{1.9} may of course be singular, that is, some (or even all) of
the $E_m$'s may coincide. In fact, the class of rational KdV solutions
as discussed by Adler and Moser \cite{am78} (see also \cite{ohm88},
\cite{om93}) arises in precisely this manner (with all $E_m=0$).
Similarly, the class of soliton solutions and more generally,
solitons relative to an algebro-geometric background potential, as
described, for instance, in
\cite{de78},
\cite{dt79},
\cite{DMN76},
\cite{gss91},
\cite{gs95},
\cite{gw93},
 \cite{Mc79}, \cite{mw97}, \cite{MEKL95}, App.~A,
\cite{Pr98} results in $n$ pairs of coinciding $E_m$'s, that is,
$\{E_m\}_{m=0,\dots,2n}=\{E_0, E_1=E_2,\dots,E_{2n-1}=E_{2n}\}$ for an
appropriate enumeration of the $E_m$'s.

Finally, a quick description of the content of this paper.
Section~\ref{KdV} is devoted to the KdV case and starts by summarizing
the basic formalism for the KdV hierarchy and its algebro-geometric
stationary solutions. Subsequently, we review Darboux transformations
$V\mapsto\widetilde V$ and finally derive the precise connection between
$\calK_n$ and $\widetilde \calK_{\tilde n}$ in
Theorem~\ref{theorem2.3}. Section~\ref{AKNS} then presents the
analogous results for the AKNS hierarchy. Besides of being an 
important hierarchy of soliton evolution equations, the AKNS
hierarchy allows us to study hyperelliptic curves without a branch
point at infinity as opposed to the KdV case, which necessarily leads
to hyperelliptic curves branched at infinity. The corresponding
Theorem~\ref{theorem3.3}, detailing the connection between $\calK_n$
and $\widetilde
\calK_{\tilde n}$ in the AKNS case, to the best of our knowledge,
appears to be without precedent.

\section{The stationary KdV hierarchy} \lb{KdV}

In this section we study B\"acklund transformations of the KdV 
hierarchy.  Our principal aim is to prove a result by Ehlers and 
Kn\"orrer \cite{ek82} on hyperelliptic curves and Darboux 
transformations 
by entirely elementary means and at the same time offer a more 
detailed
treatment (cf.~Theorem~\ref{theorem2.3}).

We start by introducing a polynomial recursion formalism following 
Al'ber \cite{al79}, \cite{al81} (see also \cite{di91}, Ch.\ 12, 
\cite{gd75}, \cite{gd79}) presented 
in detail in \cite{grt96}, \cite{gw93} (see also \cite{GH98},
\cite{gh99}, \cite{gw96}).

Suppose $V\colon\bbC\mapsto\bbCinf$, with $\bbCinf=\bbC\cup\{\infty\}$, 
is meromorphic and 
consider the Schr\"odinger operator
\begin{equation}
L = - \frac{d^2}{dx^2} + V(x), \quad x\in\bbC. \lb{2.1}
\end{equation}
Introducing $\{f_j\}_{j\in\bbN_0}$, with  $\bbN_0=\bbN\cup\{0\}$, 
recursively by
\begin{equation}
f_0 = 1,\quad f_{j,x} = - \frac14 f_{j-1, xxx} + Vf_{j-1,x}
+ \frac12 V_x f_{j-1},\quad j\in\bbN,
\lb{2.2}
\end{equation}
one finds explicitly,
\begin{align}
f_0 & =1, \no \\
f_1 & = \tfrac12 V + c_1,  \no \\
f_2 &= - \tfrac18 V_{xx} +\tfrac38 V^2 +
 c_1\tfrac12 V + c_2, \quad \text{etc.}, \lb{2.3} 
\end{align}
where $\{c_\ell\}_{\ell\in\bbN}\subset\bbC$ denote integration 
constants. The ${f}_{j}$ are well-known to be differential
polynomials in 
$V$ (see, e.g., \cite{e83}).  Given $V$ and $f_{j}$ one defines 
differential expressions $P_{2n+1}$ of order $2n+1$,
\begin{equation}
P_{2n+1}  = \sum_{j=0}^n \bigg( f_{n-j}(x)
\frac{d}{dx}-\frac12 f_{n-j,x} (x) \bigg) L^{j},
\quad n\in\bbN_0, 
\lb{2.6}
\end{equation}
and one verifies
\begin{equation}
[P_{2n+1},L] = 2f_{n+1,x}, \quad n\in\bbN_0,
\lb{2.7}
\end{equation}
with $[\dott,\dott]$ the commutator.  The 
stationary KdV hierarchy is then defined in terms of the stationary 
Lax relations
\begin{equation}
[P_{2n+1}, L] =2f_{n+1,x}=0,\quad n\in\bbN_0.
\lb{2.8}
\end{equation}
Explicitly, one finds
\begin{align} 
n&=0: \quad V_x =0, \no \\
n&=1: \quad \tfrac14 V_{xxx} -\tfrac32 VV_x - c_1V_x=0,\quad 
\text{etc.} 
\lb{2.9}
\end{align}
$V(x)$ is called an {\it algebro-geometric} KdV {\it potential} if 
it satisfies 
one (and hence infinitely many) of the equations in the 
stationary KdV 
hierarchy \eqref{2.8}.

Introducing the fundamental polynomial $F_{n}(z,x)$ of degree $n$ 
in $z$,
\begin{equation}
F_n (z,x)  = \sum_{j=0}^n f_{n-j}(x) z^j, \lb{2.10}
\end{equation}
equations \eqref{2.2} and \eqref{2.8}, that is, $f_{n+1,x}(x)=0$, 
$x\in\bbC$, imply
\begin{equation}
F_{n,xxx} -4 (V-z) F_{n,x} -2V_x F_n =0. \lb{2.11}
\end{equation}
Multiplying \eqref{2.11} by $F_{n}$ and integrating yields
\begin{equation}
\frac12 F_{n,xx}(z,x) F_n(z,x) - \frac14 F_{n,x}(z,x)^2 -
(V(x)-z) F_n(z,x)^2 = R_{2n+1}(z),
\lb{2.12}
\end{equation}
where the integration constant $R_{2n+1}(z)$ is a monic 
polynomial in 
$z$ of degree $2n+1$, and hence  of the form
\begin{equation}
R_{2n+1} (z) = \prod_{m=0}^{2n} (z-E_m),
\quad \{ E_m\}_{m=0,\dots,2n} \subset \bbC.
\lb{2.13}
\end{equation}

Introducing the algebraic eigenspace,
\begin{equation}
\ker (L-z)=\left\{ \psi\colon \bbC\mapsto\bbCinf
\text{  meromorphic}\mid (L -z)\psi=0 \right\}, \quad z\in\bbC
\lb{2.14}
\end{equation}
one verifies
\begin{equation}
P_{2n+1}\big|_{\ker(L-z)} =
\bigg(F_n (z,x) \frac{d}{dx} -\frac12F_{n,x}(z,x)\bigg)
\bigg|_{\ker(L-z)}.
\lb{2.15}
\end{equation}
Moreover, a celebrated result by Burchnall--Chaundy \cite{bc23}, 
\cite{bc28}, \cite{bc32} (see also \cite{GG91}, \cite{Pr96},
\cite{Pr98}, \cite{SW85}, \cite{Wi85})  then yields
\begin{equation}
P_{2n+1}^2+R_{2n+1}(L)=0. \lb{2.16}
\end{equation}
Equation \eqref{2.16} naturally leads to the hyperelliptic  curve 
$\dot\calK_{n}$ defined by
\begin{equation}
\dot\calK_{n}\colon \calF_n(z,y)= y^2-R_{2n+1}(z)=0, \quad
R_{2n+1}(z) = \prod_{m=0}^{2n} (z-E_m).
\label{2.17}
\end{equation}
The one-point compactification of $\dot\calK_{n}$ by joining 
$P_{\infty}$, the point at infinity, is then denoted by 
$\calK_{n}$.  
A general point $P\in\calK_{n}\backslash\{P_{\infty}\}$ will be 
denoted by $P=(z,y)$, where
\begin{equation}
\calF_n(z,y)=0. \label{2.17a}
\end{equation}
Moreover, we define the involution $*$ on  $\calK_{n}$ by
\begin{equation}
*\colon\calK_{n}\mapsto\calK_{n}, \quad P=(z,y)\mapsto 
P^{*}=(z,-y), \quad 
P_{\infty}^{*}=P_{\infty}. \label{2.18}
\end{equation}

Introducing the polynomial $H_{n+1}$ of degree $n+1$ in $z$,
\begin{equation}
H_{n+1}(z,x)=\f12F_{n,xx}(z,x)+(z-V(x))F_{n}(z,x),\label{2.19}
\end{equation}
\eqref{2.12} implies
\begin{equation}
R_{2n+1}(z)+\f14 F_{n,x}(z,x)^2 =F_{n}(z,x)H_{n+1}(z,x)	
	\label{2.20}
\end{equation}
and we may define the following fundamental meromorphic function 
$\phi(P,x)$ on $\calK_{n}$,
\begin{subequations}\lb{2.21}
\begin{align}
\phi (P,x) &= \frac{iy(P) +\frac12 F_{n,x} (z, x)}{F_n (z, x)} 
\lb{2.21a}\\
 & = \frac{-H_{n+1} (z,x)}{iy(P) -\frac12 F_{n,x}(z, x)}, \quad  
P  = (z,y)\in\calK_{n}, \; x\in\bbC, 
\lb{2.21b} 
\end{align}
where $y(P)$ denotes the meromorphic function on $\calK_{n}$ 
obtained 
upon solving $y^2=R_{2n+1}(z)$ with $P=(z,y)$.

\end{subequations}
Introducing the stationary Baker--Akhiezer function
$\psi(P, x, x_0)$ on $\calK_n\backslash \{ P_\infty \}$ by
\begin{equation}
\psi(P,x,x_0) =\exp \bigg( \int_{x_0}^x  dx' \, 
\phi(P,x')\bigg), \quad  
P  = (z,y)\in\calK_{n}\backslash\{P_{\infty}\},\,
x,x_0 \in\bbC,
\lb{2.22}
\end{equation}
choosing a smooth non-selfintersecting path from $x_{0}$ to $x$ 
which avoids singularities of $\phi (P,x)$, one easily verifies from 
\eqref{2.11}, \eqref{2.12}, \eqref{2.15}, \eqref{2.21}, and 
\eqref{2.22} the following 
result.

\begin{lemma}\lb{lemma2.1}(see, e.g., \cite{gh99}, \cite{grt96})
Suppose $f_{n+1,x}(x)=0$ and let $P= (z,y) \in \calK_n 
\backslash \{P_\infty\}$,
$x,x_0 \in \bbC$.  Then
$\phi(P,x)$ satisfies
the Riccati-type equation
\begin{equation}
 \phi_x (P,x) + \phi(P,x)^2 = V(x) -z, \lb{2.23}
\end{equation}
and
\begin{align}
 \phi(P,x) \phi (P^*, x)&= H_{n+1} (z,x) 
/ F_n (z,x),\lb{2.24}\\
\phi (P,x) + \phi (P^*, x)&= F_{n,x} (z,x) 
/ F_n (z,x),\lb{2.25}\\
 \phi(P,x)-\phi (P^*, x)&=2iy(P)/F_n (z,x), 
\lb{2.26}\\
\phi(P,x)&=iz^{-n}y(P)+\Oh(\abs{z}^{-1/2}) 
\text{  as $P=(z,y)\to P_{\infty}$}. \lb{2.27}
\end{align}
$\psi (P,x,x_0)$ satisfies 
\begin{equation}
 (L-z(P)) \psi(P,\dott,x_0) =0, 
 \quad (P_{n+1}-iy(P)) \psi(P,\dott,x_0) =0,
\lb{2.28} 
\end{equation}
and
\begin{align}
\psi (P,x,x_0) \psi (P^*, x, x_0)
&= F_n (z,x) / F_n(z,x_0),
\lb{2.29}\\
\psi_x (P,x,x_0) \psi_x (P^*, x, x_0)
&= H_{n+1} (z,x)/ F_n (z,x_0),
\lb{2.30}\\
\psi(P,x,x_0)\psi_x(P^*,x,x_0)&
+\psi(P^*,x,x_0)\psi_x(P,x,x_0) \no \\
&\qquad =F_{n,x}(z,x)/F_n(z,x_0), \lb{2.30a} \\
W(\psi (P,\dott,x_0), \psi (P^*, \dott, x_0))&=
-2iy(P)/F_{n}(z,x_0), 
\lb{2.31}
\end{align}
\begin{equation}
 \psi (P,x,x_0) = \exp
\big(iz^{-n}y(P) (x-x_0)\big)(1+\Oh(\abs{z}^{-1/2})) 
\text{  as $P=(z,y)\to P_{\infty}$}.\lb{2.32}
\end{equation}
\end{lemma}

(Here 
$W(f,g)(x)=f(x)g'(x)-f'(x)g(x)$ denotes the Wronskian 
of $f$ and $g$.)

In addition, we introduce the formal diagonal Green's function 
$G(P,x,x)$ associated with the differential expression $L$ by
\begin{subequations}\lb{2.33}
\begin{align}
G(P,x,x)&=\f{\psi (P,x,x_0) \psi (P^*, x, x_0)}
{W(\psi (P,\dott,x_0), \psi (P^*, \dott, x_0))} \lb{2.33a}\\
&=\frac{iF_n(z,x)}{2y(P)}, \quad
P=(z,y)\in\calK_n\backslash\{P_\infty\}, \; x\in\bbC,
\lb{2.33b}
\end{align}
\end{subequations}
using $\phi=\psi_{x}/\psi$, \eqref{2.26}, and \eqref{2.29}.    
Equations \eqref{2.12} 
and \eqref{2.33b} then yield the universal 
equation
\begin{multline}
-2G''(P,x,x) G(P,x,x)
 +G'(P,x,x)^2 + 4(V(x) -z) G(P,x,x)^2 =1, \\
P=(z,y)\in\calK_n\backslash\{P_\infty\}.
\lb{2.34}
\end{multline}
These observations complete our review of the stationary KdV 
hierarchy.  The time-dependent KdV hierarchy could readily be 
defined 
at this point but since it is not essential for our purpose 
we resist 
the temptation to do so (see, e.g., \cite{gh99}, \cite{grt96} for 
the time-dependent 
formalism).

Next we turn to the Darboux transformations in connection with 
$L=-d^2/dx^2+V$.  Define
\begin{multline}
\psi(P,x,x_{0},\sigma)= \begin{cases}
\f12(1+\sigma)\psi(P,x,x_{0})+\f12(1-\sigma)\psi(P^*,x,x_{0}) & 
\text{for $\sigma\in \bbC$}, \\
\psi(P,x,x_{0})-\psi(P^*,x,x_{0}) &\text{for $\sigma=\infty$},
\end{cases} \\
P\in \calK_{n}\backslash\{P_{\infty}\}, \lb{2.35}
\end{multline}
pick $Q_{0}=(z_0,y_0)\in \calK_{n}\backslash\{P_{\infty}\}$, and 
introduce the 
differential expressions
\begin{equation}
A_{\sigma}(Q_{0})=\f{d}{dx}+\phi(Q_{0},x,\sigma), \quad
A_{\sigma}^+(Q_{0})=-\f{d}{dx}+\phi(Q_{0},x,\sigma), \quad 
\sigma\in\bbCinf, \lb{2.36}
\end{equation}
where 
\begin{equation}
\phi(P,x,\sigma)=\psi_{x}(P,x,x_{0},\sigma)/\psi(P,x,x_{0},\sigma),
\quad P\in \calK_{n}\backslash\{P_{\infty}\}, \, 
\sigma\in\bbCinf. \lb{2.37}
\end{equation}
One verifies (cf.\ \eqref{2.23})
\begin{equation}
L=A_{\sigma}(Q_{0})A_{\sigma}^+(Q_{0})+z_{0}=
-\f{d^2}{dx^2}+V, 
\lb{2.38}
\end{equation}
with
\begin{equation}
V(x)=\phi(Q_{0},x,\sigma)^2+\phi_{x}(Q_{0},x,\sigma)+z_{0}, 
\lb{2.39}
\end{equation}
independent of the choice of $\sigma\in\bbCinf$.  
Interchanging the order of the differential expressions 
$A_{\sigma}(Q_{0})$ and $A_{\sigma}^+(Q_{0})$ in 
\eqref{2.38} then 
yields
\begin{equation}
\widetilde L_{\sigma}(Q_{0})=A_{\sigma}^+(Q_{0})
A_{\sigma}(Q_{0})+z_{0}
=-\f{d^2}{dx^2}+\widetilde V_{\sigma}(x,Q_{0}), \lb{2.40}
\end{equation}
with
\begin{align}
\widetilde V_{\sigma}(x,Q_{0})&=\phi(Q_{0},x,\sigma)^2-
\phi_{x}(Q_{0},x,\sigma)+z_{0}\no \\
&=V(x)-2(\ln(\psi(Q_{0},x,x_{0},\sigma)))_{xx}, \quad
\sigma\in\bbCinf.\lb{2.41}
\end{align}
The transformation
\begin{equation}
V(x)\mapsto \widetilde V_{\sigma}(x,Q_{0}), \quad
Q_{0}\in \calK_{n}\backslash\{P_{\infty}\}, \, 
\sigma\in\bbCinf \lb{2.42}
\end{equation}
is usually called the Darboux transformation (also Crum-Darboux 
transformation or single commutation method) and goes back  at 
least to 
Jacobi  \cite{j46} and Darboux \cite{d82}. While we only 
aim at its 
properties from an algebraic point of view, its analytic 
properties in 
connection with spectral deformations (isospectral and 
non-isospectral 
ones) have received enormous attention in the context of spectral 
theory (especially, regarding the insertion of eigenvalues into 
spectral gaps), inverse spectral theory, and  B\"acklund 
tranformations for the (time-dependent) KdV hierarchy.  A complete 
bibliography in this context being impossible, we just refer to 
\cite{AV94}, \cite{c55}, \cite{de78}, \cite{dt79},
\cite{ekalf82}, Ch.~4, \cite{ek82},
\cite{EF85}, \cite{fm76}, \cite{fn81}, \cite{g93}, \cite{gss91},  
\cite{gst96}, \cite{gs95}, \cite{gw93}, \cite{mc85}, \cite{mc86}, 
\cite{mc87}, \cite{sc78}, \cite{VS93} and the 
extensive literature therein. From a historical point of view it is
very interesting to note that Drach \cite{Dr18}, \cite{Dr19},
\cite{Dr19a} in his 1919 studies of Darboux transformations 
 (being a student of Darboux) not only introduced a set of
nonlinear differential equations for $V,$ which today can be
identified with the stationary KdV hierarchy, but also studied the
effect of Darboux transformations on the underlying hyperelliptic
curve. As a consequence, he seems to have been the first to
explicitly establish the connection between integrable systems and
spectral theory. For modern treatments of this connection see, for
instance, \cite{BBEIM94}, Ch.~3 and \cite{gh99}. 

Before analyzing the Darboux tranformation \eqref{2.42} in some 
detail, we 
briefly return to the formal diagonal Green's function $G(P,x,x)$ 
in 
\eqref{2.33}.  In view of the possibility of linear combinations 
as 
displayed in the definition of $\psi(P,x,x_{0},\sigma)$ in 
\eqref{2.35}, and our 
lack of natural boundary conditions (resp.\ $L^2$-conditions) in 
connection with the (possibly singular) differential expression 
$L$, 
our definition of $G(P,x,x)$ in \eqref{2.33} appears to be highly 
arbitrary.  However, the following considerations will show that 
our choice in 
\eqref{2.33} is the unique one that yields a bounded Green's 
function as 
$P\to P_{\infty}$ (for $x\in\bbC\backslash\{x_{0}\}$ fixed).  
In fact,
\begin{equation}
G(P,x,x)=\f{i z^n}{2 y(P)}+\Oh(\abs{z}^{-1}) \text{  as 
$P=(z,y)\to 
P_{\infty}$}. \lb{2.43}
\end{equation}
Introducing
\begin{multline}
H(P,x,x,x_{0},\sigma_{1},\sigma_{2})=
\f{\psi(P,x,x_{0},\sigma_{1})\psi(P,x,x_{0},\sigma_{2})}
{W(\psi(P,\dott,x_{0},\sigma_{1}),\psi(P,\dott,x_{0},
\sigma_{2}))}, 
\\ \sigma_{1}, \sigma_{2}\in\bbCinf, \, 
\sigma_{1}\neq \sigma_{2}, \lb{2.44}
\end{multline}
one infers from Lemma~\ref{lemma2.1}, \eqref{2.32}, and 
\eqref{2.35}
\begin{align}
&H(P,x,x,x_{0},\sigma_{1},\sigma_{2}) \no \\
& \quad=\f{i(1+\sigma_{1})(1+\sigma_{2})z^n}{4(\sigma_{1}-
\sigma_{2})y(P)}
\exp(2iz^{-n} y(P) (x-x_{0}))(1+\Oh(\abs{z}^{-1/2})) \no \\
& \qquad + \f{i (1-\sigma_{1})(1-\sigma_{2})z^n}{4(\sigma_{1}-
\sigma_{2})y(P)}
\exp(-2iz^{-n} y(P) (x-x_{0}))(1+\Oh(\abs{z}^{-1/2})) 
\lb{2.45} \\
& \qquad+\f{i(1-\sigma_{1}\sigma_{2})z^n}{2(\sigma_{1}-
\sigma_{2})y(P)} 
+\Oh(\abs{z}^{-1/2})  \text{  as $P=(z,y)\to P_{\infty}$} \no
\end{align}
for $\sigma_{1}, \sigma_{2}\in\bbC$, $\sigma_{1}\neq\sigma_{2}$, 
and 
similarly if $\sigma_{1}=\infty$, $\sigma_{2}\in\bbC$, or  
$\sigma_{1}\in\bbC$, $\sigma_{2}=\infty$.  Thus, 
$H(P,x,x,x_{0},\sigma_{1},\sigma_{2})$ is uniformly bounded 
for $P$ in 
a neighborhood of $P_{\infty}$ and $x\in\bbC\backslash\{x_{0}\}$ 
fixed, if and only if
\begin{equation}
\sigma_{1}=-\sigma_{2}\in\{-1,1\}. \lb{2.46}
\end{equation}
Since $\psi(P,x,x_{0},-\sigma)=\psi(P^{*},x,x_{0},\sigma)$, 
this shows 
that our choice for $G(P,x,x)$ in \eqref{2.32} yields the unique 
diagonal 
Green's function of $L$ which is bounded for $P$ in a neighborhood 
of $P_{\infty}$ for fixed $x\in\bbC\backslash\{x_{0}\}$. In 
addition, $H(P,x,x,x_{0},\sigma_{1},\sigma_{2})$ is independent of 
$x_{0}$ if and only if \eqref{2.46} holds.

Next, assuming that $\psi\in\ker(L-z)$, 
\begin{equation}
L\psi(z)=z\psi(z), \lb{2.47}
\end{equation}
one infers 
$A_{\sigma}^+(Q_{0})\psi(z)\in\ker(\widetilde 
L_{\sigma}(Q_{0})-z)$,
\begin{equation}
\widetilde L_{\sigma}(Q_{0})(A_{\sigma}^+(Q_{0})\psi(z))
=zA_{\sigma}^+(Q_{0})\psi(z), \lb{2.48}
\end{equation}
and
\begin{multline}
W(A_{\sigma}^+(Q_{0})\psi_{1}(z),
A_{\sigma}^+(Q_{0})\psi_{2}(z))
=(z-z_{0})W(\psi_{1}(z),\psi_{2}(z)), \\
\psi_{1}(z),\psi_{2}(z)\in\ker(L-z).\lb{2.49}
\end{multline}
Since
\begin{multline}
(A_{\sigma}^+(Q_{0})\psi(P,\dott,x_{0}))(x)
=\big(\phi(Q_{0},x,\sigma)-\phi(P,x)\big)\psi(P,x,x_{0}), \\
P\in\calK_{n}\backslash\{Q_{0},P_{\infty}\}, \lb{2.50}
\end{multline}
we define
\begin{align}
\tilde\psi_{\sigma}(P,x,x_{0},Q_{0}) &=
(A_{\sigma}^+(Q_{0})\psi(P,\dott,x_{0}))(x) \no \\
&=\big(\phi(Q_{0},x,\sigma)-\phi(P,x)\big)
\psi(P,x,x_{0}), \lb{2.51} \\
& \hspace*{1.5cm}  P\in\calK_{n}\backslash\{Q_{0},
P_{\infty}\}, \,
\sigma\in\bbCinf. \no
\end{align}
Then
\begin{multline}
(\widetilde L_{\sigma}(Q_{0})\tilde 
\psi_{\sigma}(P,\dott,x_{0},Q_{0}))(x)
=z\tilde\psi_{\sigma}(P,x,x_{0},Q_{0}), \\
P=(z,y)\in\calK_{n}\backslash\{Q_{0},P_{\infty}\},\lb{2.52}
\end{multline}
and we define in analogy to \eqref{2.33a} the diagonal Green's 
function 
$\widetilde G_{\sigma}(P,x,x,Q_{0})$ of $\widetilde 
L_{\sigma}(Q_{0})$ by
\begin{multline}
\widetilde G_{\sigma}(P,x,x,Q_{0})=
\f{\tilde\psi_{\sigma}(P,x,x_{0},Q_{0})
\tilde\psi_{\sigma}(P^{*},x,x_{0},Q_{0})}
{W(\tilde\psi_{\sigma}(P,\dott,x_{0},Q_{0}),
\tilde\psi_{\sigma}(P^{*},\dott,x_{0},Q_{0}))}, \\
P=(z,y)\in\calK_{n}\backslash\{Q_{0},P_{\infty}\}. \lb{2.53}
\end{multline}

\begin{lemma}\lb{lemma2.2} Assume $f_{n+1,x}(x)=0$ and let 
$Q_{0}=(z_{0},y_{0})\in\calK_{n}\backslash\{P_{\infty}\}$, 
$P=(z,y)\in\calK_{n}\backslash\{Q_{0},P_{\infty}\}$, 
$\sigma\in\bbCinf$. Then the diagonal Green's function 
$\widetilde G_{\sigma}(P,x,x,Q_{0})$ in \eqref{2.34} explicitly 
reads
\begin{subequations} \lb{2.54}
\begin{align}
\widetilde G_{\sigma}(P,x,x,Q_{0})&=
\f{H_{n+1}(z,x)+\phi(Q_{0},x,\sigma)^2F_{n}(z,x)
-\phi(Q_{0},x,\sigma)F_{n,x}(z,x)}{-2i(z-z_{0})y(P)} \lb{2.54a} \\
&=\f{(\phi(P,x)-\phi(Q_{0},x,\sigma))(\phi(P^{*},x)-
\phi(Q_{0},x,\sigma))
F_{n}(z,x)}{-2i(z-z_{0})y(P)} \lb{2.54b}\\
&=\f{i \widetilde F_{\sigma,\tilde n}(z,x)}{2\tilde y(P)}, 
\lb{2.54c}
\end{align}
\end{subequations}
where $\tilde y(P)$ denotes the meromorphic solution on 
$\dot{\widetilde\calK}_{\sigma,\tilde n}(Q_{0})$ obtained 
upon solving 
$y^2=\widetilde R_{\sigma,2\tilde n+1}(z)$, $P=(z,y)$ for some 
polynomial $\widetilde R_{\sigma,2\tilde n+1}(z)$ of degree 
$2\tilde 
n+1\in\bbN_{0}$ (cf.\ \eqref{2.17} and \eqref{2.57}) and
$\widetilde F_{\sigma,\tilde n}(z,x)$ denotes a polynomial with 
respect to $z$ of degree $\tilde n,$ with 
$0\leq\tilde n\leq n+1$.  In
particular, the Darboux  transformation 
\eqref{2.42}, $V(x)\mapsto\widetilde V_{\sigma}(x,Q_{0})$ maps 
the class 
of  algebro-geometric KdV potentials into itself.
\end{lemma}
\begin{proof}
\eqref{2.54a} and \eqref{2.54b} follow upon use of 
$\phi(P,x)=\psi_{x}(P,x,x_{0})/\psi(P,x,x_{0})$, \eqref{2.24}, 
\eqref{2.25}, 
\eqref{2.29}--\eqref{2.31}, and \eqref{2.49}.  Since the 
numerator in 
\eqref{2.54a} is 
a polynomial in $z$, and
\begin{equation}
\widetilde G_{\sigma}(P,x,x,Q_{0})=\f{iz^n}{2y(P)}+
\Oh(\abs{z}^{-1})
\text{  as $P=(z,y)\to P_{\infty}$} \lb{2.55}
\end{equation}
again by \eqref{2.54a}, one concludes \eqref{2.54c} and 
 $0\leq \tilde n\leq n+1$. By inspection,  
$\widetilde F_{\sigma,\tilde n}(z,x)$ 
satisfies equation \eqref{2.12} with $V(x)$ replaced  by 
$\widetilde V_{\sigma}(x,Q_{0})$, $n$ by $\tilde n$, and 
$R_{2n+1}(z)$ 
by $\widetilde R_{\sigma,2\tilde n +1}(z)$.  As a consequence, 
$V(x)$ being
an  algebro-geometric KdV potential implies that $\widetilde 
V_{\sigma}(x,Q_{0})$ is one as well.
\end{proof}

The following theorem, the principal result of this section, will 
clarify the dependence of $\tilde n=\tilde n(n,Q_{0},\sigma)$ on 
its variables. This result was originally derived using an entirely
different  algebro-geometric approach by Ehlers and Kn\"orrer
\cite{ek82}.

\begin{theorem}\lb{theorem2.3}
Suppose $f_{n+1,x}(x)=0$ and let 
$Q_{0}=(z_{0},y_{0})\in\calK_{n}\backslash\{P_{\infty}\}$, 
$n\in\bbN_{0}$, $\sigma\in\bbCinf$, and $\tilde 
n=\tilde n(n,Q_{0},\sigma)$ as in \eqref{2.54c}.  Then
\begin{equation}
\tilde n(n,Q_{0},\sigma)=\begin{cases}
n+1 & \text{for $\sigma\in\bbCinf\backslash\{-1,1\}$ and 
$y_{0}\neq 0$}, \\
n+1 & \text{for $\sigma=\infty$ and $y_{0}= 0$}, \\
n & \text{for $\sigma\in\{-1,1\}$ and $y_{0}\neq 0$}, \\
n & \text{for $\sigma\in\bbC$, $y_{0}=0$, and 
$R_{2n+1,z}(z_{0})\neq 0$}, \\ 
n-1 & \text{for $\sigma\in\bbC$, $y_{0}=0$, and 
$R_{2n+1,z}(z_{0})=0, n\in\bbN$}, 
\end{cases} \lb{2.56}
\end{equation}
and hence the hyperelliptic curve 
$\dot{\widetilde\calK}_{\sigma,\tilde n}(Q_{0})$ 
associated with $\widetilde V_{\sigma}(x,Q_{0})$ is of the type
\begin{equation}
\dot{\widetilde\calK}_{\sigma,\tilde n}(Q_0)\colon 
\widetilde \calF_{\sigma,\tilde n}(z,y,Q_{0})=y^2-
\widetilde R_{\sigma,2 \tilde n+1}(z,Q_{0})=0, \lb{2.56a}
\end{equation}
with
\begin{align}
&\widetilde R_{\sigma,2 \tilde n+1}(z,Q_{0}) \no \\
& =\begin{cases}
(z-z_{0})^2 R_{2n+1}(z) & \text{for 
$\sigma\in\bbCinf\backslash\{-1,1\}$ and 
$y_{0}\neq 0$}, \\
(z-z_{0})^2 R_{2n+1}(z) & \text{for $\sigma=\infty$ and 
$y_{0}= 0$}, \\
R_{2n+1}(z) & \text{for $\sigma\in\{-1,1\}$ and 
$y_{0}\neq 0$}, \\ 
R_{2n+1}(z) & \text{for $\sigma\in\bbC$, $y_{0}=0$, and 
$R_{2n+1,z}(z_{0})\neq 0$}, \\ 
(z-z_{0})^{-2} R_{2n+1}(z) & \text{for $\sigma\in\bbC$, 
$y_{0}=0$, and 
$R_{2n+1,z}(z_{0})=0, n\in\bbN$}.
\end{cases} \lb{2.57} 
\end{align}
Here
\begin{equation}
R_{2n+1}(z) =\prod_{m=0}^{2n}(z-E_{m}).
\end{equation}
\end{theorem}
\begin{proof}  Our starting point will be \eqref{2.54b} and 
a careful 
case distinction taking into account whether or not $Q_{0}$ is a 
branch point, and distinguishing the cases 
$\sigma\in\bbC\backslash\{-1,1\}$, $\sigma\in\{-1,1\}$, and 
$\sigma=\infty$.

Case (i).  $\sigma\in\bbCinf\backslash\{-1,1\}$ and 
$y_{0}\neq 0$:  One 
computes from \eqref{2.35} and \eqref{2.37},
\begin{equation}
\phi(Q_{0},x,\sigma)=\begin{cases}
\f{(1+\sigma)\psi_{x}(Q_{0},x,x_{0})+
(1-\sigma)\psi_{x}(Q_{0}^{*},x,x_{0})}
{(1+\sigma)\psi(Q_{0},x,x_{0})+(1-\sigma)\psi(Q_{0}^{*},x,x_{0})}
&\text{for $\sigma\in\bbC\backslash\{-1,1\}$}, \\[2mm]
\f{\psi_{x}(Q_{0},x,x_{0})-\psi_{x}(Q_{0}^{*},x,x_{0})}
{\psi(Q_{0},x,x_{0})-\psi(Q_{0}^{*},x,x_{0})}
&\text{for $\sigma=\infty$},
\end{cases} \lb{2.58}
\end{equation}
and upon comparison with $\phi(Q_{0},x)\neq\phi(Q_{0}^{*},x)$,
\begin{equation}
\phi(Q_{0},x)=\f{\psi_{x}(Q_{0},x,x_{0})}{\psi(Q_{0},x,x_{0})}, 
\quad
\phi(Q_{0}^{*},x)=\f{\psi_{x}(Q_{0}^{*},x,x_{0})}
{\psi(Q_{0}^{*},x,x_{0})},\lb{2.59}
\end{equation}
one concludes that no cancellations can occur in \eqref{2.54b}, 
proving $\tilde n(n,Q_{0},\sigma)=n+1$ and the first 
statement in 
\eqref{2.57}.

Case (ii).  $\sigma=\infty$ and $y_{0}=0$: Combining 
\eqref{2.21a}, 
\eqref{2.22}, \eqref{2.26}, and \eqref{2.35} one computes
\begin{align}
&\phi(Q_{0},x,\infty)=\lim_{P\to Q_{0}}\phi(P,x,\infty) \no\\
&=\lim_{P\to Q_{0}}\Bigg(
\f{\phi(P,x)\exp\big(\int_{x_{0}}^xdx'\,\phi(P,x')\big)-
\phi(P^{*},x)\exp\big(\int_{x_{0}}^xdx'\,\phi(P^{*},x')\big)}
{\exp\big(\int_{x_{0}}^xdx'\,\phi(P,x')\big)-
\exp\big(\int_{x_{0}}^xdx'\,\phi(P^{*},x')\big)}\Bigg)	\no \\
&=\phi(Q_{0},x)\no \\
&+\lim_{P\to Q_{0}}\Bigg(\f{\phi(P,x)-\phi(P^{*},x)}
{\exp\big(\int_{x_{0}}^xdx'\,\phi(P,x')\big)-
\exp\big(\int_{x_{0}}^xdx'\,\phi(P^{*},x')\big)}\times \no \\
& \hspace*{5cm} \times \exp\bigg(\int_{x_{0}}^x
dx'\,\phi(P^{*},x')\bigg)\Bigg)
\no
\\ &=\phi(Q_{0},x)+\exp\bigg(\int_{x_{0}}^xdx'\,
\phi(Q_{0},x')\bigg)\times \no \\
&\qquad\qquad\times\lim_{P\to Q_{0}}\Bigg(\f{2iy(P)/F_{n}(z,x)}
{\exp\big(iy(P)\int_{x_{0}}^x\f{dx'}{F_{n}(z,x')}\big)
-\exp\big(-iy(P)\int_{x_{0}}^x\f{dx'}{F_{n}(z,x')}\big)}\times 
\no \\
&\hspace*{6.15cm} \times\exp\bigg(-\f12\int_{x_{0}}^xdx'\, 
\f{F_{n,x}(z,x')}{F_{n}(z,x')}\bigg)\Bigg)	\no \\
&=\phi(Q_{0},x)\no \\
&\qquad+\psi(Q_{0},x,x_{0})\f{1}{F_{n}(z_{0},x)
\psi(Q_{0},x,x_{0})}
\lim_{P\to Q_{0}}\bigg(\f{2iy(P)}{2iy(P)\int_{x_{0}}^x\f{dx'}
{F_{n}(z,x')}
+\Oh(y(P)^2)}\bigg)	\no \\
&=\phi(Q_{0},x)+\bigg(F_{n}(z_{0},x)\int_{x_{0}}^x\f{dx'}
{F_{n}(z,x')}\bigg)^{-1},
\quad x\in\bbC\backslash\{x_{0}\},  \lb{2.60}
\end{align}
using $\lim_{P\to Q_{0}} y(P)=y(Q_{0})=y_{0}=0$.   From
\begin{equation}
\phi(Q_{0},x)=\f12 \f{F_{n,x}(z_{0},x)}{F_{n}(z_{0},x)}	\lb{2.61}
\end{equation}
one concludes again that no cancellations can occur in 
\eqref{2.54b}.  
Thus $\tilde n(n,Q_{0},\infty)=n+1$ and the second statement in 
\eqref{2.57} holds.

The remainder of the proof requires a more refined argument, 
the basis 
of which will be derived next.  First, replacing $V(x)$ by 
$V(x)-z_{0}$, we may assume without loss of generality that 
$z_{0}=0$ 
in the following.  Moreover, from this point on we exclude the 
trivial case $V(x)=0,$ $x\in\bbC$ (the case $V(x)=E_{0}$  
will be studied in Example~\ref{example2.5} below).  Writing
\begin{equation}
y(z)^2=R_{2n+1}(z)\underset{z\to 0}{=} y_{0}^2+
\tilde y_{1}z+\tilde 
y_{2}z^2+\Oh(z^3), \lb{2.62}	
\end{equation}
a comparison of the powers $z^0$ and $z^1$ in \eqref{2.12} 
yields 
\begin{equation}
2f_{n,xx}f_{n}=f_{n,x}^2+4Vf_{n}^2+4y_{0}^2	
	\lb{2.63}	
\end{equation}
and
\begin{equation}
f_{n-1,xx}f_{n}+f_{n,xx}f_{n-1}-f_{n,x}f_{n-1,x}
-4Vf_{n}f_{n-1}+2Vf_{n}^2-2\tilde y_{1}=0.	
		\lb{2.64}	
\end{equation}
Inserting \eqref{2.63} into \eqref{2.64}, a little algebra 
proves the 
basic identity
\begin{equation}
	f_{n}^2(f_{n-1}/f_{n})_{xx}+f_{n,x}f_{n}(f_{n-1}/f_{n})_{x}
	+2Vf_{n}^2+4y_{0}^2(f_{n-1}/f_{n})-2\tilde y_{1}=0.
	\lb{2.65}	
\end{equation}

Case (iii). $\sigma\in\{-1,1\}$ and $y_{0}\neq 0$: Then 
\eqref{2.35} 
yields
\begin{equation}
	\phi(Q_{0},x,1)=\phi(Q_{0},x), \quad \phi(Q_{0},x,-1)=
\phi(Q_{0}^{*},x),
		\lb{2.66}	
\end{equation}
with $\phi(Q_{0},x)\neq\phi(Q_{0}^{*},x)$ since $y_{0}\neq 0$. In 
this case there is a cancellation in \eqref{2.54b}.  For instance, 
choosing $\sigma=1$ one computes from \eqref{2.10} and 
\eqref{2.21a},
\begin{align}
\phi(P,x)-\phi(Q_{0},x,&1)=\phi(P,x)-\phi(Q_{0},x) \no \\
&\underset{P\to 
Q_{0}}{=}\f{i(y(P)-y_{0})}{f_{n}(x)}-iy_{0}\f{f_{n-1}(x)}
{f_{n}(x)^2}z+
\f12 \bigg(\f{f_{n-1}(x)}{f_{n}(x)}\bigg)_{x}z+\Oh(z^2)  \no \\
&\underset{P\to Q_{0}}{=}c_{1}(x)z+\Oh(z^2) \lb{2.67}
\end{align}
since
\begin{equation}
y(P)-y_{0}\underset{P\to Q_{0}}{=}y_{1}z+\Oh(z^2), \quad \tilde 
y_{1}=2y_{0}y_{1}.	\lb{2.68}	
\end{equation}
It remains to show that $c_{1}(x)$ does not vanish identically in 
$x\in\bbC$.  Arguing by contradiction we assume
\begin{equation}
	0=c_{1}(x)=\f{iy_{1}}{f_{n}(x)}-
	\f{iy_{0}}{f_{n}(x)}\f{f_{n-1}(x)}{f_{n}(x)}
	+\f12\bigg(\f{f_{n-1}(x)}{f_{n}(x)} \bigg)_{x}, \quad 
x\in\bbC.\lb{2.69}	
\end{equation}
Differentiating \eqref{2.69} with respect to $x$ and inserting 
the ensuing expression for $(f_{n-1}/f_{n})_{xx}$ and the one for 
$(f_{n-1}/f_{n})_{x}$ from \eqref{2.69} into \eqref{2.65} then 
results in the contradiction
\begin{equation}
	0=2V(x)f_{n}(x)^2, \quad x\in\bbC.\lb{2.70}	
\end{equation}
Moreover, since
\begin{equation}
\phi(P^{*},x)-\phi(Q_{0},x,x,1)=\phi(P^{*},x)-\phi(Q_{0},x)
\underset{P\to 
Q_{0}}{=}-\f{2iy_{0}}{f_{n}(x)}+\Oh(z),\lb{2.71}	
\end{equation}
one concludes that precisely one factor of $z$ cancels in 
\eqref{2.54b}.  Hence $\tilde n(n,Q_{0},1)=n$ and the third 
relation 
in \eqref{2.57} holds.  The case $\sigma=-1$ in treated 
analogously.

Case (iv). $\sigma\in\bbC$, $y_{0}=0$, and $R_{2n+1,z}(0)\neq 0$:  
Taking into account that $\phi(Q_{0},x,\sigma)=\phi(Q_{0},x)$ 
(using
\eqref{2.37} and $Q_0=Q_0^*$) is 
independent of $\sigma\in\bbC$, \eqref{2.10} and \eqref{2.21a} 
yield
\begin{equation}
	\big(\phi(P,x)-\phi(Q_{0},x) \big) \big(\phi(P^{*},x)-
\phi(Q_{0},x)   \big)
\underset{P\to Q_{0}}{=}\f{y_{1}^2 z}{f_{n}(x)^2}+\Oh(z^2)	
\lb{2.72}	
\end{equation}
since
\begin{equation}
	y(P)\underset{P\to Q_{0}}{=}y_{1}z^{1/2}+\Oh(z^{3/2}), \quad 
	y_{1}=\bigg(\prod_{E_{m}\neq 0}E_{m} \bigg)^{1/2}.
		\lb{2.73}	
\end{equation}
Thus we infer again that precisely one factor of $z$ cancels in 
\eqref{2.54b}.  Hence $\tilde n(n,Q_{0},\sigma)=n$ 
and the fourth relation in \eqref{2.57} is proved.

Case (v). $\sigma\in\bbC$, $y_{0}=\tilde y_{1}=0$, and $\tilde 
y_{2}\neq 0$ (cf.\ \eqref{2.62}):  One calculates as in 
\eqref{2.72},
\begin{align}
\big(\phi(P,x)&-\phi(Q_{0},x) \big) \big(\phi(P^{*},x)-
\phi(Q_{0},x)   
\big) \no \\
& \quad\underset{P\to Q_{0}}{=} \bigg(\f{y_{1}^2}{f_{n}(x)^2}
+\f14 \bigg(\bigg(\f{f_{n-1}(x)}{f_{n}(x)}\bigg)_{x} \bigg)^2 
\bigg)z^2+\Oh(z^3) \no \\
& \quad\underset{P\to Q_{0}}{=}c_{2}(x)z^2++\Oh(z^3)\lb{2.74}
\end{align}
since
\begin{equation}
	y(P)\underset{P\to Q_{0}}{=}y_{1}z+\Oh(z^{2}), \quad 
	y_{1}=\bigg(\prod_{E_{m}\neq 0}E_{m} \bigg)^{1/2}.
		\lb{2.75}	
\end{equation}
Next we show that $c_{2}(x)$ does not vanish identically in 
$x\in\bbC$.  Arguing again by contradiction we suppose that
\begin{equation}
0=c_{2}(x)=\f{y_{1}^2}{f_{n}(x)^2}
+\f{1}{4} \bigg(\bigg(\f{f_{n-1}(x)}{f_{n}(x)}\bigg)_{x} \bigg)^2, 
\quad x\in\bbC.\lb{2.76}	
\end{equation}
Thus
\begin{equation}
\bigg(\f{f_{n-1}}{f_{n}}\bigg)_{x}=\f{C}{f_{n}}\lb{2.77}	
\end{equation}
for some constant $C\in\bbC$.  Insertion of \eqref{2.77} and its 
$x$-derivative into \eqref{2.65} then again yields the 
contradiction
\begin{equation}
	0=2V(x)f_{n}(x)^2, \quad x\in\bbC.\lb{2.78}	
\end{equation}
Hence $\tilde n(n,Q_{0},\sigma)=n-1$ and the last 
relation in \eqref{2.57} holds in this case.

Case (vi). $\sigma\in\bbC$, $y_{0}=\tilde y_{1}=\tilde y_{2}=0$ 
(cf.\ \eqref{2.62}):  As in \eqref{2.74} one obtains
\begin{equation}
\big(\phi(P,x)-\phi(Q_{0},x) \big) \big(\phi(P^{*},x)-
\phi(Q_{0},x) \big) 
\underset{P\to Q_{0}}{=}\f{1}{4} \bigg(\bigg(\f{f_{n-1}(x)}
{f_{n}(x)}\bigg)_{x} 
\bigg)^2 z^2+\Oh(z^3)\lb{2.79}	
\end{equation}
since
\begin{equation}
	y(P)\underset{P\to Q_{0}}{=}\Oh(z^{3/2}).\lb{2.80}
\end{equation}
The remainder of the proof of case (vi) is now a special case of 
case (v) (with $y_{1}=C=0$) and one concludes again that 
$\tilde n(n,Q_{0},\sigma)=n-1$.
\end{proof}
We can summarize the previous theorem in the following table.

\bigskip
\begin{center}
\begin{tabular}{|l|c||c|c|r|}\hline
 & &$\sigma\in\bbC\backslash\{-1,1\}$ & $\sigma\in\{-1,1\}$	& 
 $\sigma=\infty$\\\hline\hline
 $y_{0}\neq 0$&& $n+1$ & $n$ & \\\cline{1-4}
 $y_{0}= 0$& $\tilde y_{1}\neq 0$ & \multicolumn{2}{c|}{$n$} 
 &  $n+1$  \\ \cline{2-4}
 & $\tilde y_{1}= 0$ & \multicolumn{2}{|c|}{$n-1$}  &  \\\hline
\end{tabular}
\medskip
\begin{tabl} \lb{ta1} The table shows the value of the 
arithmetic genus $\tilde n$  associated with the Darboux 
transformation.  Here $R_{2n+1}(z)=
y_{0}^2+\tilde y_{1}(z-z_{0})+
\Oh((z-z_{0})^2)$ as $z\to z_{0}$.
\end{tabl}
\end{center}
\bigskip

These results show, in particular, that Darboux 
transformations do 
not change the local structure of the original curve 
$y^2=R_{2n+1}(z)$, except, of course, near the point $Q_0$.

\begin{remark} \lb{remark2.4}
Theorem~\ref{theorem2.3} was first derived by purely 
algebro-geometric 
means by Ehlers and Kn\"orrer \cite{ek82} in 1982.  An 
elementary but rather lengthy derivation of
Theorem~\ref{theorem2.3}  (focusing on the case when $\tilde
n(n,\sigma)=n-1$) was provided by  Ohmiya \cite{ohm97} in 1997 
(based on two other papers 
\cite{ohm95}, \cite{om93}).  The current proof seems to 
be the only elementary and short one available at this point.  
As mentioned in the paragraph following \eqref{2.42}, Drach 
\cite{Dr18}, \cite{Dr19}, \cite{Dr19a} appears to have been the 
first to study particular aspects (the case $\tilde n=n+1$) of
Theorem~\ref{theorem2.3} around 1919. Moreover, it seems  
worthwhile 
to point out that the case $\sigma=\infty$ and $y_{0}=0$, 
which leads to  $\tilde n(n,Q_{0},\sigma)=n+1$, necessarily 
constructs an algebro-geometric KdV potential 
$\widetilde V_{\infty}(x,Q_{0})$ 
singular at $x=x_{0}$ (cf.\ \eqref{2.60}).  In fact, the class of 
rational algebro-geometric solutions constructed by Adler 
and Moser 
\cite{am78} arises exactly in this manner. 
\end{remark} 

\begin{remark} \lb{remark2.4a}
The results described in Theorem~\ref{theorem2.3} are not 
confined to hyperelliptic curves $\calK_n$ of finite 
(arithmetic) genus $n.$ In fact, upon shifting the emphasis from 
$F_n(z,x)$ to $G(P,x,x),$ the results in Theorem~\ref{theorem2.3} 
extend to certain classes of transcendental hyperelliptic curves of
infinite (arithmetic) genus $\calK_\infty$ (including those
associated with periodic potentials $V$). More details will be 
presented elsewhere. 
\end{remark}

We conclude with the following elementary illustration.

\begin{example}\lb{example2.5}  The case $n=0$.
\begin{align}
y(P)^2&=R_{1}(z)=z-E_{0},  \no \\
F_{0}(z,x)&=1, \quad H_{1}(z,x)=z-E_{0}, \quad V(x)=E_{0}, 
\no \\
\phi(P,x)&=iy(P), \quad \psi(P,x,x_{0})=\exp(iy(P)(x-x_{0})),
\no \\
G(P,x,x)&=\f{i}{2y(P)},\no \\
\phi(P,x,\sigma)&=\begin{cases}
iy(P)\f{(1+\sigma)\exp(iy(P)(x-x_{0}))-(1-\sigma)
\exp(-iy(P)(x-x_{0}))}
{(1+\sigma)\exp(iy(P)(x-x_{0}))+(1-\sigma)\exp(-iy(P)(x-x_{0}))}, 
& \sigma\in\bbC, \\[2mm]
iy(P)\f{\exp(iy(P)(x-x_{0}))-\exp(-iy(P)(x-x_{0}))}
{\exp(iy(P)(x-x_{0}))+\exp(-iy(P)(x-x_{0}))}, 
& \sigma=\infty,
\end{cases} \no \\
\phi((E_{0},0),x,\sigma)&=\begin{cases}
0, & \sigma\in\bbC, \\
(x-x_{0})^{-1}, &\sigma=\infty.
\end{cases}
\end{align}
More generally, the case $V(x)=E_{0}$ can be associated with any 
curve $y(P)^2=R_{2n+1}(z)$ since $f_{n,x}(x)=0$, 
$n\in\bbN_{0}$ in this special case. 
\end{example}

\section{The stationary AKNS hierarchy} \lb{AKNS}

This section is devoted to Darboux (gauge) transformations 
for the 
AKNS hierarchy.  In particular, we derive the KdV analogs of 
Section~\ref{KdV} and hence determine the effect of gauge tranformations 
on 
hyperelliptic AKNS curves in the spirit of our approach to the KdV 
result of Ehlers and Kn\"orrer \cite{ek82}.

We start by introducing a polynomial recursion for the AKNS 
hierarchy 
following the derivation in \cite{gr98}.  Suppose 
$p,q\colon\bbC\mapsto\bbCinf$ are meromorphic and introduce the 
Dirac-type 
differential expression
\begin{equation}
D=\begin{pmatrix}\f{d}{dx} & -q \\ p & -\f{d}{dx}  \end{pmatrix}, 
\quad x\in\bbC. \lb{3.1}	
\end{equation}
Introducing $\{f_{j}(x)\}_{j\in\bbN_{0}}$, 
$\{g_{j}(x)\}_{j\in\bbN_{0}}$, 
and $\{h_{j}(x)\}_{j\in\bbN_{0}}$  recursively
by
\begin{align}
 f_0(x)&=-iq(x),\quad g_0(x)=1,\quad h_0(x)=ip(x), \no \\
 f_{j}(x)&= \f{i}{2} f_{j-1,x}(x) - i q(x) g_{j}(x), \quad
 j\in\bbN, \no \\ 
g_{j,x}(x)&= p(x)f_{j-1}(x) +
 q(x)h_{j-1}(x),\quad j\in\bbN, \label{3.2} \\
 h_{j}(x)&= -\f{i}{2} h_{j-1,x}(x) + i p(x) g_{j}(x), \quad
 j\in\bbN. \no
\end{align}
Explicitly, one computes
\begin{align}
f_0& = -iq,\quad f_1=\tfrac 12 q_x+c_1 (-iq),\no \\
f_2& = \tfrac{i}{4} q_{xx}-
      \tfrac{i}{2}pq^2+ c_1 \tfrac12 q_x
+c_2 (-iq), \no \\
g_0& = 1,\quad g_1=c_1,\quad g_2=\tfrac12 pq+c_2,\no \\
g_3& =  -\tfrac{i}{4}(p_{_x}q - pq_x) 
+ c_1\tfrac12 pq 
+ c_3 ,
\label{3.3}\\ 
h_0& =i p,\quad h_1=\tfrac12 p_{_x}+ c_1 ip, \no \\
h_2&=-\tfrac{i}{4} p_{_{xx}}+\tfrac{i}{2}p{^2}q
+ c_1 \tfrac12
p_{_x}+c_2i p, \quad \text{etc.}, \no
\end{align}
where $\{c_j\}_{j\in\bbN_0}\subset\bbC$ are integration 
constants. The coefficients ${f}_j$, ${g}_j$, ${h}_j$ are 
well-known 
to be differential polynomials in $p$ and $q$ (see, e.g., 
\cite{gh99}).  Given $p$, $q$ and $f_{j}$, $g_{j}$, and 
$h_{j}$ one 
defines the matrix-valued differential expression of order $n+1$,
\begin{equation}
E_{n+1}=i \sum_{j=0}^{n+1}
\begin{pmatrix}   
-g_{n+1-j} & f_{n-j} \\ -h_{n-j} & g_{n+1-j}
 \end{pmatrix}D^{j}, \quad n\in \bbN_0,\quad f_{-1}=h_{-1}=0,
\label{3.6}
\end{equation}
and verifies 
\begin{equation}
[E_{n+1},D]= \begin{pmatrix}
 0 & -2if_{n+1} \\ 2 i h_{n+1} & 0 \end{pmatrix}, \quad n\in 
\bbN_0.
\label{3.7}
\end{equation}
The stationary AKNS hierarchy is then defined by the 
stationary Lax 
relations
\begin{equation}
	[E_{n+1},D]=0, \quad n\in\bbN, \text{  that is,  } 
	f_{n+1}=h_{n+1}=0, \quad n\in\bbN.
	\label{3.8}
\end{equation}
Explicitly, one finds
\begin{align}
  n&=0: \quad\begin{pmatrix}
    - p_{_x} + c_1(-2 i p)  \\
   - q_x + c_1(2 i q) 
  \end{pmatrix}= 0, \no \\
 n&=1: \quad \begin{pmatrix} 
    \f{i}{2} p_{_{xx}}-i p{^2}q +
      c_1( - p_{_x})+c_2(-2 i p) \\[2mm]
    -\f{i}{2} q_{xx}+i pq^2+ c_1(-q_x)+c_2(2 i q)
  \end{pmatrix} = 0,  \text{  etc}. \lb{3.9}
\end{align}
$(p(x),q(x))$ are called {\it algebro-geometric} AKNS 
{\it potentials} if 
they satisfy one (and hence infinitely many) of the equations 
of the 
stationary AKNS hierarchy \eqref{3.8}.

Introducing the polynomials $F_{n}(z,x)$, $G_{n}(z,x)$, and 
$H_{n}(z,x)$ 
with respect to $z$,
\begin{align}
F_n(z,x)&= \sum_{j=0}^{n}f_{n-j}(x)z^{j}, \no \\
G_{n+1}(z,x)&= \sum_{j=0}^{n+1}g_{n+1-j}(x)z^j, 
\label{3.10}\\
H_n(z,x)&= \sum_{j=0}^{n}h_{n-j}(x)z^j, \no
\end{align}
equations \eqref{3.2} and \eqref{3.8}, that is, 
$f_{n+1}(x)=h_{n+1}(x)=0$, $x\in\bbC$, imply
\begin{equation}
F_{n,x}=-2izF_{n}+2qG_{n+1}, \quad G_{n+1,x}=pF_{n}+qH_{n}, 
\quad
H_{n,x}=2izH_{n}+2pG_{n+1}.
\label{3.11}
\end{equation}
Equations \eqref{3.11} yield
\begin{equation}
G_{n+1}(z,x)^2- F_n(z,x)H_n(z,x)=R_{2n+2}(z),
\label{3.12}
\end{equation}
where  $R_{2n+2}(z)$ is a monic 
polynomial in $z$ of degree $2n+2$ and hence of the form
\begin{equation}
R_{2n+2}(z)= \prod_{m=0}^{2n+1} (z-E_m),\quad
\{E_m\}_{m=0,\dots,2n+1}
\subset \bbC.
\label{3.13}
\end{equation}
Moreover, \eqref{3.11} and \eqref{3.12} imply
\begin{align}
&F_n(z,x)F_{n,xx}(z,x)-\f{q_x(x)}{q(x)}F_n(z,x)F_{n,x}(z,x)-
\f12F_{n,x}(z,x)^2 \no \\
&+\bigg(2z^2-2iz
\f{q_x(x)}{q(x)}-2p(x)q(x)\bigg)F_n(z,x)^2 =-2 q(x)^2 R_{2n+2}(z),
\lb{3.14} \\
&H_n(z,x)H_{n,xx}(z,x)-\f{p_x(x)}{p(x)}H_n(z,x)H_{n,x}(z,x)
-\f12H_{n,x}(z,x)^2 \no \\
&+\bigg(2z^2+2iz
\f{p_x(x)}{p(x)}-2p(x)q(x)\bigg)H_n(z,x)^2  =-2 p(x)^2 R_{2n+2}(z).
\lb{3.15}
\end{align}
Introducing the algebraic eigenspace
\begin{equation}
\ker(D-z)=\left\{ 
\Psi =\begin{pmatrix}{\psi}_1 \\ 
{\psi}_2\end{pmatrix}\colon\bbC\mapsto(\bbCinf)^2\text{ meromorphic}
 \mid (D-z)\Psi=0 \right\}, \quad z \in \bbC,
\label{3.16}
\end{equation}
one verifies
\begin{equation}
E_{n+1}\Big{\vert}_{\ker(D-z)}= i
\begin{pmatrix}
   -G_{n+1}(z,x) & F_{n}(z,x) \\ 
   -H_{n}(z,x)  & G_{n+1}(z,x)
\end{pmatrix} 
\bigg{\vert}_{\ker(D-z)}.
\label{3.17}
\end{equation}
Moreover, the analog of the Burchnall--Chaundy polynomial for 
the KdV 
case in Section~\ref{KdV} now reads
\begin{equation}
E_{n+1}^2+R_{2n+2}(D)=0. \lb{3.18}	
\end{equation}
Equation \eqref{3.18} naturally leads to the hyperelliptic curve 
$\dot\calK_{n}$ defined by 
\begin{equation}
\dot\calK_{n}\colon\calF_n(z,y)=y^2-R_{2n+2}(z)=0, \quad
  R_{2n+2}(z) = \prod_{m=0}^{2n+1} (z-E_m), \,\, z\in \bbC.
\label{3.19}
\end{equation}
The compactification of $\dot\calK_{n}$, by joining 
$\{\Pinfmin,\Pinfplus\}$, the points at infinity, is then denoted 
$\calK_{n}$.  As in Section~\ref{KdV} we denote points 
$P\in\calK_{n}\backslash\{\Pinfmin,\Pinfplus\}$  by $P=(z,y)$, 
where 
$\calF_n(z,y)=0$.
Moreover the involution $*$ on $\calK_{n}$ is defined by
\begin{equation}
*\colon\calK_{n}\mapsto\calK_{n}, \quad P=(z,y)\mapsto 
P^{*}=(z,-y), \quad 
P_{\infty_{\mp}}^{*}=P_{\infty_{\pm}}. \label{3.20aa}
\end{equation}
Because of \eqref{3.12} we may define the following fundamental 
meromorphic function $\phi(P,x)$ on $\calK_{n}$
\begin{subequations}\label{3.20}
\begin{align}
\phi(P,x)&=\f{y(P)+G_{n+1}(z,x)}{F_n(z,x)}\lb{3.20a}\\
&=\f{-H_n(z,x)}
{y(P)-G_{n+1}(z,x)}, \quad
P=(z,y) \in \calK_n, \, x\in\bbC,
\label{3.20b}
\end{align}
\end{subequations}
where $y(P)$ denotes the meromorphic function on $\calK_n$ 
obtained 
upon solving $y^2=R_{2n+2}(z)$ denoting $P=(z,y)$.  The 
associated Baker--Akhiezer 
vector $\Psi(P,x,x_{0})$ on 
$\calK_n\backslash\{\Pinfmin,\Pinfplus\}$ 
is then defined by
\begin{align}
\psi_{1}(P,x,x_{0})&=\exp\bigg(\int_{x_{0}}^x 
dx'\,\big(-iz+q(x')\phi(P,x')\big) \bigg), \no \\
\psi_{2}(P,x,x_{0})&=\phi(P,x)\psi_{1}(P,x,x_{0}), \no \\
\Psi(P,x,x_{0})&=\begin{pmatrix}\psi_{1}(P,x,x_{0})\\ 
\psi_{2}(P,x,x_{0})
\end{pmatrix}, \quad P=(z,y) \in 
\calK_n\backslash\{\Pinfmin,\Pinfplus\},\,x,x_{0}\in\bbC,\lb{3.21}
\end{align}
choosing a small non-selfintersecting path from $x_{0}$ to $x$ 
avoiding singularities of $q(x)$ and $\phi(P,x)$.  The analog of 
Lemma~\ref{lemma2.1} then reads as follows.

\begin{lemma}\lb{lemma3.1} (see, e.g., \cite{gh99}, \cite{gr98})
Suppose $f_{n+1}=h_{n+1}=0$ and let $P=(z,y) \in 
\calK_n\backslash\{\Pinfmin,\Pinfplus\}$, $x,x_{0}\in\bbC$.  
Then $\phi(P,x)$ satisfies the Riccati-type equation
\begin{equation}
\phi_x(P,x)+q(x)\phi(P,x)^2-2iz\phi(P,x)=p(x), \label{3.22}
\end{equation}
and
\begin{align}
 \phi(P,x)\phi(P^*,x)&=H_{n} (z, x)/F_{n} (z, x), 
\label{3.23} \\
 \phi(P,x)+\phi(P^*,x)&=2G_{n+1} (z, x)/F_{n} (z, x), 
\label{3.24} \\
 \phi(P,x)-\phi(P^*,x)&= 2 y(P)/F_{n} (z, x). \label{3.25}
\end{align}
Moreover, if $p(x)q(x)\neq 0$,
\begin{equation}
\phi(P,x)=\Oh(z^{\mp 1}) \text{  as $P=(z,x)\to 
P_{\infty_{\pm}}$}.\label{3.26}	
\end{equation}
$\Psi(P,x,x_{0})$ satisfies
\begin{equation}
(D-z(P))\Psi(P,\dott,x_{0})=0,\quad (E_{n+1}-iy(P))
\Psi(P,\dott,x_{0})=0,
\lb{3.27}
\end{equation}
and
\begin{align}
 \psi_1(P,x,x_0)\psi_1(P^*,x,x_0)&=
  F_{n} (z, x)/F_{n} (z, x_0), \label{3.28} \\
 \psi_2(P,x,x_0)\psi_2(P^*,x,x_0)&=
  H_{n} (z, x)/F_{n} (z, x_0), \label{3.29}\\
\psi_1(P,x,x_0)\psi_2(P^*,x,x_0)&+
    \psi_1(P^*,x,x_0)\psi_2(P,x,x_0) \no \\
   &\qquad=2 G_{n+1}(z,x)/F_{n}(z,x_0),\label{3.30} \\
W(\Psi(P,\dott,x_{0}),\Psi(P^{*},\dott,x_{0}))&
=-2y(P)/F_{n}(z,x_0). 
\label{3.31}  
\end{align}
Moreover, if $p(x)q(x)\neq 0$, $q(x_{0})\neq 0$, then
\begin{align}
 \psi_1(P,x,x_0)&=\exp(z^{-n}y(P)(x-x_{0}))(1+\Oh(z^{-1})), 
\no \\
 \psi_2(P,x,x_0)&=\exp(z^{-n}y(P)(x-x_{0}))\Oh(z^{\mp1})
 \text{  as $P=(z,y)\to P_{\infty_{\pm}}$}.\label{3.32}  
\end{align}
\end{lemma}

(Here $W(F,G)=f_{1}g_{2}-f_{2}g_{1}$ denotes the Wronskian of 
$F=\big(\begin{smallmatrix} f_{1} \\ f_{2}\end{smallmatrix}
\big)$ and
$G=\big(\begin{smallmatrix} g_{1} 
\\ g_{2}\end{smallmatrix}\big)$.)

Since the diagonal elements $G_{\ell,\ell}(P,x,x')$, $\ell=1,2$ of 
the $2\times 2$ Green's matrix $G(P,x,x')$, $x\neq x'$ associated 
with $D$ are discontinuous as $x\to x'$ in contrast to the 
off-diagonal elements $G_{\ell,\ell'}(P,x,x')$, $\ell\neq\ell'$, 
$\ell,\ell'=1,2$, we consider
\begin{subequations}\lb{3.33}
\begin{align}
G_{1,2}(P,x,x)&=-i\f{\psi_{1}(P,x,x_{0})\psi_{1}(P^{*},x,x_{0})}
{W(\Psi(P,\dott,x_{0}),\Psi(P^{*},\dott,x_{0}))}
\lb{3.33a} \\
&=\f{iF_{n}(z,x)}{2y(P)},\quad 
P=(z,y) \in 
\calK_n\backslash\{\Pinfmin,\Pinfplus\}, \, x\in\bbC, \lb{3.33b} 
\end{align}
\end{subequations}
using $\phi=\psi_{2}/\psi_{1}$, \eqref{3.25}, and \eqref{3.28}.  
Equations \eqref{3.14} and \eqref{3.33b} then yield the universal 
equation
\begin{multline}
2G_{1,2}^{\prime\prime}(P,x,x)G_{1,2}(P,x,x)-2\f{q_{x}(x)}{q(x)}
G_{1,2}^{\prime}(P,x,x)G_{1,2}(P,x,x)\\
-G_{1,2}^{\prime}(P,x,x)^2 
+2\bigg(2z^2-2iz\f{q_{x}(x)}{q(x)}-2p(x)q(x)\bigg)
G_{1,2}(P,x,x)^2=q(x)^2,
\\ P=(z,y) \in 
\calK_n\backslash\{\Pinfmin,\Pinfplus\}. \lb{3.34}
\end{multline}
Similarly, we find
\begin{subequations}\lb{3.33A}
\begin{align}
G_{2,1}(P,x,x)&=-i\f{\psi_{2}(P,x,x_{0})\psi_{2}(P^{*},x,x_{0})}
{W(\Psi(P,\dott,x_{0}),\Psi(P^{*},\dott,x_{0}))}
\lb{3.33Aa} \\
&=\f{iH_{n}(z,x)}{2y(P)},\quad 
P=(z,y) \in 
\calK_n\backslash\{\Pinfmin,\Pinfplus\}, \, x\in\bbC, \lb{3.33Ab} 
\end{align}
\end{subequations}
and
\begin{multline}
G_{2,1}^{\prime\prime}(P,x,x)G_{2,1}(P,x,x)-2\f{p_x(x)}{p(x)}
G_{2,1}^{\prime}(P,x,x)G_{2,1}(P,x,x) \\
-G_{2,1}^{\prime}(P,x,x)^2+2\bigg(2 z^2+2iz
\f{p_x(x)}{p(x)}-2p(x)q(x)\bigg)G_{2,1}(P,x,x)^2=p(x)^2, \\
P=(z,y) \in 
\calK_n\backslash\{\Pinfmin,\Pinfplus\}, \, x\in\bbC. \lb{3.34A}
\end{multline}

This completes our review of the stationary AKNS hierarchy.  The 
corresponding time-dependent AKNS hierarchy can now readily be 
introduced, see, for  instance, \cite{gh99}, \cite{gr98}.

Next we turn to gauge (i.e., Darboux-type) transformations in 
connection with 
$D=i\big(\begin{smallmatrix}d/dx & -q \\ p& -d/dx 
\end{smallmatrix}\big)$.  Introducing 
\begin{equation}
	U(z,x)=
\begin{pmatrix}-iz & -q(x) \\ p(x)& iz\end{pmatrix},
	\lb{3.35}	
\end{equation}
the equation
\begin{equation}
D\Psi(z,x)=z\Psi(z,x), \quad \Psi(z,x)=\begin{pmatrix}
\psi_{1}(z,x) \\ 
\psi_{2}(z,x) \end{pmatrix} \lb{3.36}	
\end{equation}
is equivalent to
\begin{equation}
\Psi_{x}(z,x)=U(z,x)\Psi(z,x).	
\lb{3.37}	
\end{equation}
The formal gauge transformation
\begin{align}
\Psi(z,x)\mapsto \widetilde\Psi(z,x)&=\Gamma(z,x)\Psi(z,x), 
\lb{3.38} \\
U(z,x))\mapsto \widetilde U(z,x)&=\begin{pmatrix}-iz 
& -\tilde q(x) 
\\ \tilde p(x)& 
iz\end{pmatrix} \no \\
&=\Gamma(z,x)U(z,x)\Gamma(z,x)^{-1}
+\Gamma_{x}(z,x)\Gamma(z,x)^{-1}, \lb{3.39}
\end{align}
with $\Gamma(z,x)$ a $2\times 2$ matrix to be chosen later, then 
implies
\begin{equation}
\widetilde\Psi_{x}(z,x)=\widetilde U(z,x)\widetilde\Psi(z,x).	
\lb{3.39a}	
\end{equation}
Hence,
\begin{equation}
\widetilde D\widetilde\Psi(z,x)=z\widetilde\Psi(z,x), \quad 
\widetilde\Psi(z,x)=\begin{pmatrix}\tilde\psi_{1}(z,x) \\ 
\tilde\psi_{2}(z,x) \end{pmatrix}, \lb{3.40}	
\end{equation}
with
\begin{equation}
\widetilde D=i\begin{pmatrix}d/dx & -\tilde q \\ \tilde p& -d/dx 
\end{pmatrix}.
\lb{3.41}	
\end{equation}
Next, introduce
\begin{multline}
\Psi(P,x,x_{0},\sigma)=\begin{cases}
\f12(1+\sigma)\Psi(P,x,x_{0})+\f12(1-\sigma)\Psi(P^{*},x,x_{0})
& \text{for $\sigma\in\bbC$}, \\
\Psi(P,x,x_{0})-\Psi(P^{*},x,x_{0})
& \text{for $\sigma=\infty$},\end{cases} \\
P=(z,y) \in 
\calK_n\backslash\{\Pinfmin,\Pinfplus\}, \lb{3.42}	
\end{multline}
pick $Q_{0}=(z_{0},y_{0}) \in 
\calK_n\backslash\{\Pinfmin,\Pinfplus\}$, and define
\begin{equation}
\Gamma(Q_{0},x,\sigma)=\begin{pmatrix}
z-z_{0}-\f{i}2 q(x) \phi(Q_{0},x,\sigma) & \f{i}2 q(x) \\
\f{i}2\phi(Q_{0},x,\sigma) & -\f{i}2
\end{pmatrix}.\lb{3.43}	
\end{equation}
Here $\Psi(P,x,x_{0})$ is defined in \eqref{3.20} and
\begin{align}
\phi(P,x,\sigma)&=\psi_{2}(P,x,x_{0},
\sigma)/\psi_{1}(P,x,x_{0},\sigma), \lb{3.44}	\\ 
& P=(z,y) \in \calK_n\backslash\{\Pinfmin,\Pinfplus\}, \,
\sigma\in\bbCinf. \no
\end{align}
According to \eqref{3.39} one then obtains
\begin{subequations}\lb{3.45} 
\begin{align}
\widetilde D_{\sigma}(Q_{0})
&=i\begin{pmatrix}d/dx & -\tilde q_{\sigma}(Q_{0}) \\ 
\tilde p_{\sigma}(Q_{0})& -d/dx 
\end{pmatrix}, \quad \sigma\in\bbCinf, \lb{3.45a}	\\
\tilde p_{\sigma}(x,Q_{0})&=\phi(Q_{0},x,\sigma), \lb{3.45b} \\
\tilde 
q_{\sigma}(x,Q_{0})&=-2iz_{0}q(x)-q_{x}(x)+\phi(Q_{0},x,
\sigma)q(x)^2, 
\\ 
&\qquad  Q_{0}=(z_{0},y_{0}) \in 
\calK_n\backslash\{\Pinfmin,\Pinfplus\}, \, 
\sigma\in\bbCinf,\lb{3.45c}
\end{align}
\end{subequations}
utilizing the fact that $\phi(P,x,\sigma)$ satisfies the 
Riccati-type 
equation \eqref{3.22} for all $\sigma\in\bbCinf$. The gauge 
transformation (or equivalently, Darboux transformation)
\begin{equation}
(p(x),q(x))\mapsto 
(\tilde p_{\sigma}(x), \tilde q_{\sigma}(x))\lb{3.46a}	
\end{equation}
can be inferred from the results in \cite{kr92} (with a bit of 
additional work). Adding solitons (i.e., inserting eigenvalues 
into the spectrum of $D$) and its effect on the Baker-Akhiezer
vector $\Psi$ has also been studied in
\cite{Fl83}, \cite{fn81}. 

We also mention that as in Section~\ref{KdV} 
(cf.\ 
\eqref{2.43}--\eqref{2.44}), our choice of the  off-diagonal 
elements 
of the Green's matrix  of $D$ in \eqref{3.33} yields the only 
bounded Green's matrix for $P$ near $P_{\infty_{\pm}}$ and generic 
$x\in\bbC$ when comparing with more general linear combinations 
$\Psi(P,x,x_{0},\sigma)$ in \eqref{3.42} as opposed to 
$\Psi(P,x,x_{0})$ or $\Psi(P^{*},x,x_{0})$.  Introducing
\begin{multline}
\widetilde\Psi_{\sigma}(P,x,x_{0},Q_{0})=
\Gamma(Q_{0},x,\sigma)\Psi(P,x,x_{0}),
\\ P=(z,y) \in 
\calK_n\backslash\{\Pinfmin,\Pinfplus\}, \, \sigma\in\bbCinf,
\lb{3.46}	
\end{multline}
where $\Psi(P,\dott,x_{0})\in\ker(D-z)$ and 
$\Gamma(Q_{0},x,\sigma)$ 
is defined in \eqref{3.43}, one infers
\begin{equation}
\widetilde D_{\sigma}(Q_{0})\widetilde
\Psi_{\sigma}(P,x,x_{0},Q_{0})
=z\widetilde\Psi_{\sigma}(P,x,x_{0},Q_{0}).\lb{3.47}	
\end{equation}
Moreover,
\begin{align}
&W(\widetilde\Psi_{\sigma,1}(P,\dott,x_{0},Q_{0}),
\widetilde\Psi_{\sigma,2}(P,\dott,x_{0},Q_{0})) \no \\
&=-\f{i}2 (z-z_{0})W(\Psi_{1}(P,\dott,x_{0}),\Psi_{2}
(P,\dott,x_{0})),
\lb{3.48}	
\end{align}
where
\begin{multline}
\widetilde\Psi_{\sigma,j}(P,x,x_{0},Q_{0})
=\Gamma(Q_{0},x,\sigma)\Psi_{j}(P,x,x_{0}), \\
\Psi_{j}(P,\dott,x_{0})\in\ker(D-z), \, j=1,2.
\lb{3.49}	
\end{multline}
Given these facts we define in analogy to \eqref{3.33a} and 
\eqref{3.33Aa} the 
off-diagonal Green's matrix elements associated with 
$\widetilde D_\sigma (Q_0)$ by 
\begin{align}
\widetilde G_{\sigma,1,2}(P,x,x,Q_{0})
&=-i\f{\tilde\psi_{\sigma,1}(P,x,x_{0},Q_{0})
\tilde\psi_{\sigma,1}(P^{*},x,x_{0},Q_{0})}
{W(\widetilde\Psi_{\sigma}(P,\dott,x_{0},Q_{0}),
\widetilde\Psi_{\sigma}(P^{*},\dott,x_{0},Q_{0}))},\lb{3.49A}\\
\widetilde G_{\sigma,2,1}(P,x,x,Q_{0})
&=-i\f{\tilde\psi_{\sigma,2}(P,x,x_{0},Q_{0})
\tilde\psi_{\sigma,2}(P^{*},x,x_{0},Q_{0})}
{W(\widetilde\Psi_{\sigma}(P,\dott,x_{0},Q_{0}),
\widetilde\Psi_{\sigma}(P^{*},\dott,x_{0},Q_{0}))},\lb{3.50} \\
& \qquad \qquad\qquad P=(z,y) \in 
\calK_n\backslash\{Q_{0},\Pinfmin,\Pinfplus\},\,\, 
\sigma\in\bbCinf. \no
\end{align}

\begin{lemma} \lb{lemma3.2}
Assume $f_{n+1}=h_{n+1}=0$ and let $Q_{0}=(z_{0},y_{0}) \in 
\calK_n\backslash\{\Pinfmin,\Pinfplus\}$, $P=(z,y) \in 
\calK_n\backslash\{Q_{0},\Pinfmin,\Pinfplus\}$,  
$\sigma\in\bbCinf$. Then the off-diagonal elements of the 
Green's matrix of $\widetilde D_\sigma (Q_0)$ read
\begin{subequations}\lb{3.51} 
\begin{align}
&\widetilde G_{\sigma,1,2}(P,x,x,Q_{0})  \no\\
&\quad=i\bigg(\bigg(i(z-z_{0})q(x)
+\f12q(x)^2\phi(Q_{0},x,\sigma)\bigg)G_{n+1}(z,x)
-\f14q(x)^2H_{n}(z,x) \no \\
&\qquad\quad +\bigg((z-z_{0})^2-i(z-z_{0})q(x)\phi(Q_{0},x,\sigma)
-\f14q(x)^2\phi(Q_{0},x,\sigma)^2\bigg)F_{n}(z,x)
\bigg)\times \no \\
&\qquad\qquad\quad\,\, \times\big(-(z-z_{0})y(P)\big)^{-1}
\lb{3.51a} \\
&\quad=i\bigg((z-z_{0})+\f{i}2 
q(x)\big(\phi(P,x)-\phi(Q_{0},x,\sigma)\big)\bigg)\times \no \\
& \qquad\quad\times \bigg((z-z_{0}) +\f{i}2 q(x) 
\big(\phi(P^{*},x)-\phi(Q_{0},x,\sigma)\big)\bigg)F_{n}(z,x)
\big(-(z-z_{0})y(P)\big)^{-1}\lb{3.51b} \\
&\quad= \f{i\widetilde F_{\sigma,\tilde n}(z,x)}
{2\tilde y(P)} \lb{3.51c} 
\end{align}
\end{subequations}
and
\begin{subequations}\lb{3.51A}
\begin{align} 
&\widetilde G_{\sigma,2,1}(P,x,x,Q_{0})  \no\\
&\quad =\f{i}{4} \big(\phi(Q_{0},x,\sigma)^2F_{n}(z,x)
-2\phi(Q_{0},x,\sigma) G_{n+1}(z,x) +
H_{n}(z,x)  \big) \times \no \\
&\hspace*{11mm} \times \big((z-z_0)y(P)  \big)^{-1}\lb{3.51d} \\
&\quad =\f{i}4 \big(\phi(P,x)-\phi(Q_{0},x,\sigma)\big)
\big(\phi(P^{*},x)-\phi(Q_{0},x,\sigma)\big)F_n(z,x)
\big((z-z_0)y(P)  \big)^{-1} \lb{3.51e}  \\
&\quad= \f{i\widetilde H_{\sigma,\tilde n}(z,x)}
{2\tilde y(P)},\lb{3.51f} 
\end{align}
\end{subequations}
where $\tilde y(P)$ denotes the meromorphic solution on 
$\dot{\widetilde\calK}_{\sigma,\tilde n}(Q_{0})$ obtained 
upon solving 
$y^2=\widetilde R_{\sigma,2\tilde n+2}(z)$, $P=(z,y)$ for some 
polynomial $\widetilde R_{\sigma,2\tilde n+2}(z)$ of degree 
$2\tilde n+2$ and 
$\widetilde F_{\sigma,\tilde n}(z,x)$ and 
$\widetilde H_{\sigma,\tilde n}(z,x)$ denote polynomials 
in $z$ of
degree $\tilde n$, with $0\leq\tilde n\leq n+1$. 
In particular, the Darboux transformation \eqref{3.46a}, 
$(p(x),q(x))\mapsto (\tilde p_{\sigma}(x), 
\tilde q_{\sigma}(x))$ maps 
the class of algebro-geometric AKNS potentials into itself. 
\end{lemma}
\begin{proof} We present the argument for 
$\widetilde G_{\sigma,1,2}$
only, the case $\widetilde G_{\sigma,2,1}$ follows similarly.
As in Lemma~\ref{lemma3.2}, 
$\phi(P,x)=\psi_{2}(P,x,x_{0})/\psi_{1}(P,x,x_{0})$, 
\eqref{3.23}, 
\eqref{3.24}, \eqref{3.28}--\eqref{3.31}, and \eqref{3.48} imply 
equations \eqref{3.51a} and \eqref{3.51b}. Since the numerator in 
\eqref{3.51a} is a polynomial in $z$, and
\begin{equation}
\widetilde G_{\sigma,1,2}(P,x,x,Q_{0})=\f{\tilde 
q(x)z^n}{2y(P)}+\Oh(\abs{z}^{-2}) \text{  as $P=(z,y)\to 
P_{\infty_{\pm}}$} \lb{3.52}
\end{equation}
by \eqref{3.51a}, one infers \eqref{3.51c} and 
$0\leq\tilde n\leq n+1$.  Again one  verifies that 
$\widetilde F_{\sigma,\tilde n}(z,x)$ satisfies equation 
\eqref{3.14} 
with $(p(x),q(x))$ replaced by $(\tilde p_{\sigma}(x), \tilde 
q_{\sigma}(x)),$ $n$ by $\tilde n$, and $R_{2n+2}(z)$  
by 
$\widetilde R_{\sigma,2\tilde n+2}(z),$ proving that 
the Darboux
transformation 
\eqref{3.46a} leaves the class of algebro-geometric AKNS 
potentials 
invariant.
\end{proof}

The following theorem, the principal result of this section 
and in 
complete analogy to Theorem~\ref{theorem2.3}, will clarify the 
dependence of $\tilde n=\tilde n(n,Q_{0},\sigma)$ on 
its variables.

\begin{theorem} \lb{theorem3.3}
Suppose $f_{n+1}=h_{n+1}=0$, let $Q_{0}=(z_{0},y_{0}) \in 
\calK_n\backslash\{\Pinfmin,\Pinfplus\}$, $n\in\bbN_{0}$,
$\sigma\in\bbCinf$, and $\tilde n=\tilde n(n,Q_{0},\sigma)$ 
as in 
\eqref{3.51c}.  Then
\begin{equation}
\tilde n(n,Q_{0},\sigma)=\begin{cases}
n+1 & \text{for $\sigma\in\bbCinf\backslash\{-1,1\}$ and 
$y_{0}\neq 0$}, \\
n+1 & \text{for $\sigma=\infty$ and $y_{0}= 0$}, \\
n & \text{for $\sigma\in\{-1,1\}$ and $y_{0}\neq 0$}, \\
n & \text{for $\sigma\in\bbC$, $y_{0}=0$, and 
$R_{2n+2,z}(z_{0})\neq 0$}, \\ 
n-1 & \text{for $\sigma\in\bbC$, $y_{0}=0$, and 
$R_{2n+2,z}(z_{0})=0, n\in\bbN$}, 
\end{cases} \lb{3.53}
\end{equation}
and hence the hyperelliptic curve
 $\dot{\widetilde\calK}_{\sigma,\tilde n}(Q_{0})$ 
associated with $(\tilde p_{\sigma}(x,Q_{0}),\tilde 
q_{\sigma}(x,Q_{0}))$ 
is of the type
\begin{equation}
\dot{\widetilde\calK}_{\sigma,\tilde n}\colon 
\widetilde \calF_{\sigma,\tilde n}(z,y,Q_{0})=y^2-
\widetilde R_{\sigma,2 \tilde n+2}(z,Q_{0})=0, \lb{3.54a}
\end{equation}
with
\begin{align}
&\widetilde R_{\sigma,2 \tilde n+2}(z,Q_{0}) \no \\
&=\begin{cases}
(z-z_{0})^2 R_{2n+2}(z) & \text{for 
$\sigma\in\bbCinf\backslash\{-1,1\}$ and 
$y_{0}\neq 0$}, \\
(z-z_{0})^2 R_{2n+2}(z) & \text{for $\sigma=\infty$ and 
$y_{0}= 0$}, \\
R_{2n+2}(z) & \text{for $\sigma\in\{-1,1\}$ and 
$y_{0}\neq 0$}, \\
R_{2n+2}(z) & \text{for $\sigma\in\bbC$, $y_{0}=0$, and 
$R_{2n+2,z}(z_{0})\neq 0$}, \\ 
(z-z_{0})^{-2} R_{2n+2}(z) & \text{for $\sigma\in\bbC$, 
$y_{0}=0$, and 
$R_{2n+2,z}(z_{0})=0, n\in\bbN$}.
\end{cases} \lb{3.54}
\end{align}
Here
\begin{equation}
R_{2n+2}(z) =\prod_{m=0}^{2n+1}(z-E_{m}).\lb{3.54aa}
\end{equation}
\end{theorem}
\begin{proof}
The following arguments closely parallel those in the proof of 
Theorem~\ref{theorem2.3}.  Again our starting point will be 
\eqref{3.51b} and \eqref{3.51e} and a careful case distinction 
between 
$\sigma\in\bbCinf\backslash\{-1,1\}$, $\sigma\in\{-1,1\}$, 
$\sigma=\infty$, and whether or not $Q_{0}$ is a branch point.

Case (i). $\sigma\in\bbCinf\backslash\{-1,1\}$ and 
$y_{0}\neq 0$: One 
calculates using \eqref{3.42} and \eqref{3.44}, 

\begin{equation}
\phi(Q_{0},x,\sigma)=\begin{cases}
\f{(1+\sigma)\psi_{2}(Q_{0},x,x_{0})+
(1-\sigma)\psi_{2}(Q_{0}^{*},x,x_{0})}
{(1+\sigma)\psi_{1}(Q_{0},x,x_{0})+
(1-\sigma)\psi_{1}(Q_{0}^{*},x,x_{0})}
&\text{for $\sigma\in\bbC\backslash\{-1,1\}$}, \\[2mm]
\f{\psi_{2}(Q_{0},x,x_{0})-\psi_{2}(Q_{0}^{*},x,x_{0})}
{\psi_{1}(Q_{0},x,x_{0})-\psi_{1}(Q_{0}^{*},x,x_{0})}
&\text{for $\sigma=\infty$},
\end{cases} \lb{3.55}
\end{equation}
and since  
\begin{equation}
\phi(Q_{0},x)=\f{\psi_{1}(Q_{0},x,x_{0})}
{\psi_{2}(Q_{0},x,x_{0})}\neq
\phi(Q_{0}^{*},x)=\f{\psi_{1}(Q_{0}^{*},x,x_{0})}
{\psi_{2}(Q_{0}^{*},x,x_{0})},\lb{3.56}
\end{equation}
one concludes that no cancellations can occur in 
\eqref{3.51b} or 
\eqref{3.51e} and hence
$\tilde n(n,Q_{0},\sigma)=n+1$.

Case (ii). $\sigma=\infty$ and $y_{0}=0$:  Then \eqref{3.20a}, 
\eqref{3.21}, \eqref{3.25} and \eqref{3.42} imply
\begin{align}
&\phi(Q_{0},x,\infty)=\lim_{P\to Q_{0}}\phi(P,x,\infty) \no\\
&=\lim_{P\to Q_{0}}\bigg(
\bigg(\phi(P,x)\exp\bigg(\int_{x_{0}}^xdx'\,\big(-iz+q(x')
\phi(P,x')\big)\bigg) \no\\
&\qquad\qquad\qquad -\phi(P^{*},x)\exp\bigg(\int_{x_{0}}^xdx'\,
\big(-iz+q(x')
\phi(P^{*},x')\big)\bigg)\bigg)\times \no\\
&\qquad\qquad\qquad\qquad\times\bigg(\exp\bigg(\int_{x_{0}}^xdx'\,
\big(-iz+q(x')
\phi(P,x')\big)\bigg)\no\\
&\qquad\qquad\qquad\qquad\qquad -\exp\bigg(\int_{x_{0}}^xdx'\,
\big(-iz+q(x')
\phi(P^{*},x')\big)\bigg)\bigg)^{-1}\bigg)	\no \\
&=\phi(Q_{0},x)\no \\
&\quad +\lim_{P\to Q_{0}}\Bigg(\f{\phi(P,x)-\phi(P^{*},x)}
{\exp\big(\int_{x_{0}}^xdx'\,q(x')\phi(P,x')\big)-
\exp\big(\int_{x_{0}}^xdx'\,q(x')\phi(P^{*},x')\big)}\times \no \\
&\hspace*{6cm} \times
\exp\bigg(\int_{x_{0}}^xdx'\,q(x')\phi(P^{*},x')\bigg)\Bigg)	\no \\
&=\phi(Q_{0},x)+\exp\bigg(\int_{x_{0}}^xdx'\,
q(x')\phi(Q_{0},x')\bigg)\times \no \\
&\times\lim_{P\to Q_{0}}\bigg(
\bigg(\bigg(\exp\bigg(y(P)\int_{x_{0}}^xdx'\,\f{q(x')}
{F_{n}(z,x')}\bigg)
-\exp\bigg(-y(P)\int_{x_{0}}^x dx'\,\f{q(x')}{F_{n}(z,x')}\bigg)
\bigg)\times 
\no \\
&\hspace*{5.1cm} \times\exp\bigg(\int_{x_{0}}^xdx'\, 
\f{q(x')G_{n+1}(z,x')}{F_{n}(z,x')}\bigg)\bigg)^{-1}\f{2y(P)}
{F_{n}(z,x)}\bigg)	
\no \\
&=\phi(Q_{0},x)+\bigg(F_{n}(z_{0},x)\int_{x_{0}}^x dx'\,
\f{q(x')}{F_{n}(z,x')}\bigg)^{-1},
\quad x\in\bbC\backslash\{x_{0}\},  \lb{3.57}
\end{align}
using $\lim_{P\to Q_{0}} y(P)=y(Q_{0})=y_{0}=0$.  Since by 
\eqref{3.11}
\begin{equation}
\phi(Q_{0},x)=\f{G_{n+1,x}(z_{0},x)}{F_{n}(z_{0},x)}
=\f{1}{q(x)}\bigg(\f{F_{n,x}(z_{0},x)}{2F_{n}(z_{0},x)}
+iz_{0}\bigg),
\lb{3.58}
\end{equation}
one again concludes  that no cancellation occurs in 
\eqref{3.51b} or 
\eqref{3.51e},  
and hence $\tilde n(n,Q_{0},\infty)=n+1$.

The rest of the proof relies on some additional arguments 
to be discussed next.  
First, we will assume without loss of generality that 
$z_{0}=0$.  This 
can be achieved by noticing that
\begin{equation}
D=i\begin{pmatrix}d/dx & - q(x) \\  p(x)& -d/dx 
\end{pmatrix} \text{  and  }
D_{a}=i\begin{pmatrix}d/dx & - q(x)e^{-2ax} \\  p(x)e^{2ax} 
& -d/dx 
\end{pmatrix}
\lb{3.59}
\end{equation}
are related by
\begin{equation}
U D_{a}U^{-1}=D-iaI, \quad 
U=\begin{pmatrix}e^{2ax} & 0 \\ 0& e^{-2ax} \end{pmatrix}.	
\lb{3.60}
\end{equation}
Moreover, we will exclude the trivial case where $p(x)$ and 
$q(x)$ are
 proportional to 
$e^{-2ic_{1}x}$ and
$e^{2ic_{1}x},$ $x\in\bbC,$ respectively (cf.\
Example~\ref{example3.4}  below).
 We write
\begin{equation}
y(z)^2=R_{2n+2}(z)\underset{z\to 0}{=} y_{0}^2+\tilde y_{1}z
+\tilde 
y_{2}z^2+\Oh(z^3). \lb{3.61}	
\end{equation}

Case (iii). $\sigma\in\{-1,1\}$ and $y_{0}\neq 0$: Then 
\eqref{3.42} 
yields
\begin{equation}
	\phi(Q_{0},x,1)=\phi(Q_{0},x), \quad \phi(Q_{0},x,-1)
=\phi(Q_{0}^{*},x),
		\lb{3.65}	
\end{equation}
with $\phi(Q_{0},x)\neq\phi(Q_{0}^{*},x)$ since 
$y_{0}\neq 0$.  In 
this case there is a cancellation in \eqref{3.51b} and 
\eqref{3.51e}.  
Choosing $\sigma=1$ 
one computes from \eqref{3.10} and \eqref{3.20a},
\begin{equation}
\phi(P^{*},x)-\phi(Q_{0},x,1)=\phi(P^{*},x)-
\phi(Q_{0},x)\underset{P\to 
Q_{0}}{=}-\f{2y_{0}}{f_{n}(x)}+\Oh(z).\lb{3.70}	
\end{equation}
Furthermore,
\begin{align}
&\phi(P,x)-\phi(Q_{0},x,1)
=\phi(P,x)-\phi(Q_{0},x) \no \\
&\underset{P\to Q_{0}}{=}\f{y(P)-y_{0}}{f_{n}(x)}
-\bigg(y_{0}\f{f_{n-1}(x)}{f_{n}(x)^2}-
\f{g_{n}(x)}{f_{n}(x)}+\f{g_{n+1}(x)f_{n-1}(x)}
{f_{n}(x)^2}\bigg)z
+\Oh(z^2) \no \\
&\underset{P\to Q_{0}}{=}
\bigg(\f{y_{1}}{f_{n}(x)}-
\f{y_{0}f_{n-1}(x)}{f_{n}(x)^2}+\f{g_{n}(x)}{f_{n}(x)}
-\f{g_{n+1}(x)f_{n-1}(x)}{f_{n}(x)^2}\bigg)z+\Oh(z^2)\no \\
&\underset{P\to Q_{0}}{=}c_{1}(x)z+\Oh(z^2)  \lb{3.66} 
\end{align}
since
\begin{equation}
y(P)-y_{0}\underset{P\to Q_{0}}{=}y_{1}z+\Oh(z^2), \quad \tilde 
y_{1}=2y_{0}y_{1}.	\lb{3.67}	
\end{equation}
Similarly, we find
\begin{equation}
z+\f{i}{2}q(x)\big(\phi(P,x)-\phi(Q_{0},x,1)\big)
\underset{P\to
Q_{0}}{=}\bigg(1+\f{i}{2}q(x)c_1(x)\bigg)z+\Oh(z^2). \lb{3.68}
\end{equation}
It remains to show that $c_1$ does not vanish identically. 
We assume temporarily that $\widetilde G_{\sigma,1,2}$ and
$\widetilde G_{\sigma,2,1}$ have cancellations of the
same order as $z\to 0.$ Arguing by contradiction we suppose 
that $c_1$  vanishes identically. But \eqref{3.66} and 
\eqref{3.68} then
show that
$\widetilde G_{\sigma,1,2}$ and $\widetilde G_{\sigma,2,1}$ 
would have
cancellations of different order, which is a contradiction. 
We conclude that precisely one factor of $z$ cancels in 
\eqref{3.51b} and \eqref{3.51e}, and hence 
$\tilde n(n,Q_{0},1)=n$. The case $\sigma=-1$ is treated 
analogously. It remains to show that $\widetilde G_{\sigma,1,2}$ 
and $\widetilde G_{\sigma,2,1}$ necessarily have cancellations 
of the same order as $z\to 0.$ A comparison of 
$\widetilde D_\sigma=i\big(\begin{smallmatrix}d/dx 
& -\tilde q_\sigma \\ \tilde p_\sigma & -d/dx 
\end{smallmatrix}\big)$ and its formal adjoint $\widetilde 
D_\sigma^*=i\big(\begin{smallmatrix}d/dx 
& -\overline{\tilde p_\sigma} \\ \overline{\tilde q_\sigma} 
& -d/dx \end{smallmatrix}\big)$ yields the replacement of 
$(\tilde p_\sigma, \tilde q_\sigma)$ by 
$(\overline{\tilde q_\sigma}, \overline{\tilde p_\sigma})$ 
and hence the corresponding  replacements of 
$(\widetilde F_{\sigma,
\tilde n}(z,x), \widetilde G_{\sigma, \tilde n +1}(z,x), 
\widetilde H_{\sigma, \tilde n}(z,x))$ by
$\left(\overline{\widetilde F_{\sigma,
\tilde n}(\overline z,x)}, \overline{\widetilde G_{\sigma, 
\tilde n +1}(\overline z,x)}, \overline{\widetilde H_{\sigma, 
\tilde n}(\overline z,x)}\right)$ (cf.~\eqref{3.2} and 
\eqref{3.10})
and $\widetilde R_{\sigma, 2\tilde n +2}(z)$ by 
$\overline{\widetilde R_{\sigma,2\tilde n +2}(\overline z)}$ 
(cf.~the notation employed in Lemma~\ref{lemma3.2}). This fact 
has two consequences: Firstly, from relation \eqref{3.12} 
we infer
that the corresponding algebraic curves associated with 
$\widetilde D_\sigma$  and $\widetilde D_\sigma^*$ have complex
conjugate  branch points, that is, if 
$\{\widetilde E_{\sigma, m}\}_{m=0,\dots,2\tilde n +1}$ 
corresponds to $\widetilde D_\sigma,$  then 
$\left\{\overline{\widetilde E_{\sigma, m}}\right\}_{m=0,
\dots,2\tilde n +1}$
corresponds to $\widetilde D_\sigma^*,$  where $\widetilde
R_{\sigma,2\tilde n +2}(z)=\prod_{m=0}^{2\tilde n +1}
(z-\widetilde E_{\sigma,m}).$ Secondly, we infer
\begin{equation}
\widetilde G_{\sigma,2,1}(P,x,x)=\overline{ \widetilde
G_{\sigma,2,1}^* (\overline P,x,x)}, \lb{3.70a}
\end{equation}
where $P(z,y),$ $\overline P =(\overline z,\overline y),$ 
and $\widetilde G_\sigma^*(P,x,x)$ denotes the Green's 
matrix associated 
with $\widetilde D_\sigma^*.$ This shows that any 
cancellations in  
$\widetilde G_{\sigma,1,2}$ and $\widetilde G_{\sigma,2,1}$ as 
$z\to 0$ are necessarily of identical order.

Case (iv). $\sigma\in\bbC$, $y_{0}=0$, and 
$R_{2n+1,z}(0)\neq 0$:  
Using $\phi(Q_{0},x,\sigma)=\phi(Q_{0},x)$ for all 
$\sigma\in\bbC$, 
\eqref{3.10} and \eqref{3.20a} yield
\begin{align}
&\bigg(z+\f{i}2 q(x)	\big(\phi(P,x)-\phi(Q_{0},x) \big)\bigg) 
\bigg(z+\f{i}2 q(x)\big(\phi(P^{*},x)-\phi(Q_{0},x)   
\big)\bigg) \no\\
&\underset{P\to Q_{0}}{=}\f{q(x)^2y_{1}^2}{4f_{n}(x)^2}z
+\Oh(z^2)
\lb{3.71}	
\end{align}
since
\begin{equation}
	y(P)\underset{P\to Q_{0}}{=}y_{1}z^{1/2}+\Oh(z^{3/2}), \quad 
	y_{1}=\bigg(\prod_{E_{m}\neq 0}E_{m} \bigg)^{1/2}.
		\lb{3.72}	
\end{equation}
Thus  again  precisely one factor of $z$ cancels in 
\eqref{3.51b} (similarly, one factor cancels in 
\eqref{3.51e}), and 
hence $\tilde n(n,Q_{0},\sigma)=n$.

Case (v). $\sigma\in\bbC$, $y_{0}=\tilde y_{1}=0$, and $\tilde 
y_{2}\neq 0$ (cf.\ \eqref{3.61}):  One computes as in 
\eqref{3.71},
\begin{align}
&\bigg(z+\f{i}2 q(x)\big(\phi(P,x)-\phi(Q_{0},x) \big)\bigg)
\bigg(z+\f{i}2 q(x) \big(\phi(P^{*},x)-\phi(Q_{0},x)   
\big)\bigg) \no \\
& \underset{P\to Q_{0}}{=} \bigg(\f{q(x)^2y_{1}^2}{4f_{n}(x)^2}
+\bigg(1+\f{i}2 q(x)\bigg(\f{g_{n}(x)}{f_{n}(x)}
-\f{g_{n+1}(x)f_{n-1}(x)}{f_{n}(x)^2}\bigg)\bigg)^2 
\bigg)z^2+\Oh(z^3) \no \\
& \underset{P\to Q_{0}}{=}c_{2}(x)z^2+\Oh(z^3)\lb{3.73}
\end{align}
since
\begin{equation}
	y(P)\underset{P\to Q_{0}}{=}y_{1}z+\Oh(z^{2}), \quad 
	y_{1}=\bigg(\prod_{E_{m}\neq 0}E_{m} \bigg)^{1/2}.
		\lb{3.74}	
\end{equation}
Similarly, we find
\begin{align}
&\big(\phi(P,x)-\phi(Q_{0},x) \big)\big(\phi(P^{*},x)-
\phi(Q_{0},x)   
\big) \no \\
&\underset{P\to Q_{0}}{=}-\bigg(\f{y_{1}^2}
{f_{n}(x)^2}
-\bigg(\f{g_{n}(x)}{f_{n}(x)}
-\f{g_{n+1}(x)f_{n-1}(x)}{f_{n}(x)^2}\bigg)^2 
\bigg)z^2+\Oh(z^3) \no \\
&\underset{P\to Q_{0}}{=}c_3(x)z^2+\Oh(z^3). \lb{3.75}
\end{align}
We see that both $c_2(x)$ and $c_3(x)$ cannot vanish 
simultaneously,
and hence precisely a factor $z^2$ cancels in 
\eqref{3.51b} and \eqref{3.51e}. Thus $\tilde n(n,Q_{0},
\sigma)=n-1$.

Case (vi). $\sigma\in\bbC$, $y_{0}=\tilde y_{1}=\tilde y_{2}=0$ 
(cf.\ \eqref{3.61}):  In analogy to \eqref{3.73} and \eqref{3.75} 
one obtains
\begin{align}
&\bigg(z+\f{i}2 q(x)\big(\phi(P,x)-\phi(Q_{0},x) \big)\bigg)
\bigg(z+\f{i}2 q(x) \big(\phi(P^{*},x)-\phi(Q_{0},x)   
\big)\bigg) \no\\
&\underset{P\to Q_{0}}{=}\bigg(1+\f{i}2 q(x) 
\bigg(\f{g_{n}(x)}{f_{n}(x)}-\f{g_{n+1}f_{n-1}}{f_{n}^2}
\bigg)\bigg)^2z^2
+\Oh(z^3)\lb{3.79}	
\end{align}
and
\begin{align}
&\big(\phi(P,x)-\phi(Q_{0},x) \big)
\big(\phi(P^{*},x)-\phi(Q_{0},x)   \big) \no \\
&\underset{P\to Q_{0}}{=}
\bigg(\f{g_{n}(x)}{f_{n}(x)}-\f{g_{n+1}f_{n-1}}{f_{n}^2}\bigg)^2z^2
+\Oh(z^3)\lb{3.79a}	
\end{align}
respectively, since
\begin{equation}
	y(P)\underset{P\to Q_{0}}{=}\Oh(z^{3/2}).\lb{3.80}
\end{equation}
Thus this case subordinates to case (v) resulting again in
$\tilde n(n,Q_{0},\sigma)=n-1$.
\end{proof}

While Lemma~\ref{lemma3.2} has first been noted in \cite{GW98} 
(see also \cite{gw98}), Theorem~\ref{theorem3.3} appears 
to be new. We also emphasize that  Table~\ref{ta1} and
Remark~\ref{remark2.4a} apply in the present  AKNS context. 

We conclude this section with the elementary genus zero example.

\begin{example} \lb{example3.4}
The case $n=0$.
\begin{align}
	y(P)^2&=R_{2}(z)=(z-E_{0})(z-E_{1}),  \no \\
	c_{1}&=-(E_{0}+E_{1})/2, \no \\
	p(x)&=p(x_{0})e^{-2ic_{1}(x-x_{0})}, \quad 	
	q(x)=q(x_{0})e^{2ic_{1}(x-x_{0})}, \no \\
	p(x)q(x)&=(E_{0}-E_{1})^2/4, \no \\
	F_{0}(z,x)&=-iq(x), \quad G_{1}(z,x)=z+c_{1}, 
\quad H_{0}(z,x)=ip(x),\no \\
	\phi(P,x)&= \f{y(P)+z+c_{1}}{-iq(x)}=\f{ip(x)}{y(P)-z-c_{1}}, 
\no \\
	\psi_{1}&=e^{i(y(P)+c_{1})(x-x_{0})}, \quad 
	\psi_{2}=\f{y(P)+z+c_{1}}{-iq(x_{0})}e^{i
(y(P)-c_{1})(x-x_{0})}, \no \\
	\phi(P,x,\sigma)&=\f{i}{q(x)}\times \no \\
	&\hspace*{-21mm} \times\begin{cases}
\f{(1+\sigma)(y(P)+z+c_{1})\exp(iy(P)(x-x_{0}))+
(1-\sigma)(-y(P)+z+c_{1})
\exp(-iy(P)(x-x_{0}))}
{(1+\sigma)\exp(iy(P)(x-x_{0}))+(1-\sigma)
\exp(-iy(P)(x-x_{0}))}, 
& \sigma\in\bbC, \\[2mm]
\f{(y(P)+z+c_{1})\exp(iy(P)(x-x_{0}))+(y(P)-z-c_{1})
\exp(-iy(P)(x-x_{0}))}
{\exp(iy(P)(x-x_{0}))+\exp(-iy(P)(x-x_{0}))}, 
& \sigma=\infty,
\end{cases} \no \\
\phi((E_{j},0),x,\sigma)&=\f{1}{2q(x)}\begin{cases}
i(E_j+c_1), &  \sigma\in\bbC, \\
i(E_{j}+c_{1})+(x-x_{0})^{-1},& \sigma=\infty,
\end{cases} \quad j=0,1. \lb{3.81}
\end{align}
\end{example}
\vspace*{3mm}

\noindent {\bf Acknowledgments.}
We are indebted to Rudi Weikard for discussions on Darboux-type
transformations for AKNS systems.


\begin{thebibliography}{99}
%
\bi{AV94}
M.~Adler and P.~van Moerbeke, {\it Birkhoff strata, B\"acklund
transformations, and regularization of isospectral operators}, 
Adv. Math. {\bf 108}, 140--204 (1994).
%
\bibitem{am78}	M.~Adler and J.~Moser,
{\it On a class of polynomials connected with the Korteweg-de
Vries equation}, Comm. Math. Phys. {\bf 61}, 1--30 (1978).
%
\bibitem{al79} S.~I.~Al'ber, \textit{Investigation of 
equations of
Korteweg-de Vries type by the method of recurrence relations},
J. London Math. Soc. {\bf 19}, 467--480  (1979) (Russian).
%
\bibitem{al81} S.~I.~Al'ber, {\it On stationary problems for
equations of Korteweg-de
Vries type\/}, Commun.\ Pure Appl.\ Math.\
{\bf 34}, 259--272 (1981).
%
\bibitem{BBEIM94} E. D. Belokolos, A. I. Bobenko, V. Z.
Enol'skii,
A.R. Its,
and V. B. Matveev, \textit{Algebro-Geometric Approach to
Nonlinear
Integrable Equations}, Springer, Berlin, 1994.
%
\bibitem{bc23}
J.~L.~Burchnall and T.~W.~Chaundy, {\it Commutative
ordinary differential
operators\/}, Proc.\ London Math.\ Soc.\ (2),
{\bf 21}, 420--440 (1923).
%
\bibitem{bc28}
J.~L.~Burchnall and T.~W.~Chaundy, {\it Commutative
ordinary differential
operators\/}, Proc.\ Roy.\ Soc.\ London {\bf A118},
557--583 (1928).
%
\bibitem{bc32}
J.~L.~Burchnall and T.~W.~Chaundy, {\it Commutative ordinary 
differential 
operators II. The identity $P^n=Q^m$},
Proc.\ Roy.\ Soc.\ London {\bf A134}, 471--485 (1932).
%
\bibitem{c55} M.~M.~Crum,
{\it Associated Sturm-Liouville systems},
Quart. J. Math. Oxford Ser. (2)
{\bf 6}, 121--127 (1955).
%
\bibitem{d82} G.~Darboux,
{\it Sur une proposition relative aux \'{e}quations 
lin\'{e}aire},
C. R. Acad. Sci. Paris S{\'e}r. I Math. {\bf 94}, 
1456--1459 (1882).
%
\bibitem{de78} P.~A.~Deift,
{\it Applications of a commutation formula},
Duke Math. J. {\bf 45}, 267--310 (1978).
%
\bibitem{dt79}
P.~Deift and E.~Trubowitz,
{\it Inverse scattering on the line},
Comm. Pure Appl. Math. {\bf 32}, 121--251 (1979).
%
\bibitem{di91} L.~A.~Dickey, {\it Soliton Equations and
Hamiltonian Systems\/},
World Scientific, Singapore, 1991.
%
\bibitem{Dr18}
J.~Drach, \textit{Sur les groupes complexes de 
rationalit\'e et
sur l'int\'egration par quadratures,} C. R. Acad. 
Sci. Paris
{\bf 167} (1918), 743--746.
%
\bibitem{Dr19}
J.~Drach, \textit{D\'etermination des cas de
r\'eduction
de l'\'equation diff\'erentielle $d^2 y/dx^2=[\phi(x)
+h]y$}, C. R.
Acad. Sci. Paris {\bf 168} (1919), 47--50.
%
\bibitem{Dr19a}  
J.~Drach, \textit{Sur l'int\'egration par
quadratures
de l'\'equation $d^2 y/dx^2=[\phi(x)+h]y$}, C. R. Acad. Sci.
Paris {\bf 168} (1919), 337--340.
%
\bi{DMN76}
B.~A.~Dubrovin, V.~B.~Matveev, and S.~P.~Novikov,
{\it Non-linear equations
of Korteweg-de Vries type, finite-zone linear
operators, and Abelian
varieties}, Russian Math. Surv. {\bf 31:1},
59--146 (1976).
%
\bibitem{ekalf82}
M.~S.~P.~Eastham and H.~Kalf,
{\it Schr\"odinger-Type Operators with Continuous Spectra},
Pitman, Boston, 1982.
%
\bibitem{ek82} F.~Ehlers and H.~Kn\"orrer, \textit{An
algebro-geometric interpretation of the B\"acklund
transformation for the Korteweg-de Vries equation},
Comment. Math. Helv. {\bf 57}, 1--10 (1982).
%
\bibitem{e83}
G.~Eilenberger, {\it Solitons}, Springer, Berlin,
1983.
%
\bi{EF85}
N. M. Ercolani and H. Flaschka, {\it The geometry of 
the Hill equation and of the Neumann system}, Phil. 
Trans. Roy. Soc. London \textbf{A 315}, 405--422 (1985).
%
\bibitem{Fl83}
H.~Flaschka, {\it Relations between infinite-dimensional 
and finite-dimensional isospectral equations}, in 
{\it Non-linear Integrable Systems -- Classical Theory 
and Quantum Theory}, M.~Sato (ed.), World Scientific, 
Singapore, 1983, pp.~219--240. 
%
\bibitem{fm76}
H.~Flaschka and D.~W.~McLaughlin,
{\it Some comments on B\"acklund transformations, canonical
transformations, and the inverse scattering method}, 
in {\it B\"acklund Transformations, the Inverse 
Scattering Method, Solitons, and their Applications}, 
R.~M.~Miura (ed.),
Lecture Notes in Math. {\bf 515}, Springer,
Berlin, 1976, pp.~252--295.
%
\bibitem{fn81}
H.~Flaschka and A.~C.~Newell, {\it Multiphase similarity 
solutions of 
integrable evolution equations}, Physica {\bf 3D}, 
203--221 (1981).
%
\bibitem{GG91} L.~Gatto and S.~Greco, \textit{Algebraic 
curves and
differential equations: an introduction}, The Curves 
Seminar at
Queen's,
Vol. VIII (ed. by A.~V.~Geramita),  Queen's Papers 
Pure Appl.
Math.
{\bf 88}, Queen's Univ., Kingston, Ontario, Canada, 
1991, B1--B69.
%
\bibitem{gd75}  I.~M.~Gel'fand and L.~A.~Dikii, {\it Asymptotic
behaviour of the
resolvent of Sturm-Liouville equations and the
algebra of the Korteweg-de
Vries equations\/}, Russian Math.\ Surv.\
{\bf 30:5}, 77--113 (1975).
%
\bibitem{gd79} I.~M.~Gel'fand and L.~A.~Dikii, {\it Integrable
nonlinear equations and
the Liouville theorem\/}, Funct.\ Anal.\ Appl.\
{\bf 13}, 6--15 (1979).
%
\bibitem{g93}
F.~Gesztesy, {\it A complete spectral characterization of
the double commutation method},
J. Funct. Anal. {\bf 117}, 401--446 (1993).
%
\bibitem{GH98} F.~Gesztesy and H.~Holden,
{\it Dubrovin equations and integrable systems on hyperelliptic 
curves}, preprint, 1998.
%
\bibitem{gh99} F.~Gesztesy and H.~Holden,
{\it Hierarchies of Soliton Equations and Their 
Algebro-Geometric 
Solutions}, monograph in preparation.
%
\bibitem{gr98} F.~Gesztesy and R.~Ratneseelan, \textit{An
alternative approach to algebro-geometric solutions of 
the AKNS hierarchy}, Rev. Math. Phys. {\bf10}, 345--391 (1998). 
%
\bibitem{grt96} F.~Gesztesy, R.~Ratnaseelan, and G.~Teschl,
{\it The KdV hierarchy and associated trace formulas},
in {\it Recent Developments in Operator Theory and its
Applications}.
Operator Theory: Advances and Applications,
{\bf 87}, I.~Gohberg, P.~Lancaster, and P.~N.~Shivakumar (eds.),
Birkh\"auser, Basel, 1996, pp.~125--163.
%
\bibitem{gss91}
F.~Gesztesy, W.~Schweiger, and B.~Simon,
{\it Commutation methods applied to the mKdV-equation},
Trans. Amer. Math. Soc. {\bf 324}, 465--525 (1991).
%
\bibitem{gst96}
F.~Gesztesy, B.~Simon and G.~Teschl,
{\it Spectral deformation of one-dimensional Schr\"{o}dinger 
operators},
J. Analyse Math. {\bf 70}, 267--324 (1996).
%
\bibitem{gs95} F.~Gesztesy and R.~Svirsky,
{\it (m)KdV solitons on the background of
quasi-periodic finite-gap solutions},
Mem. Amer. Math. Soc. {\bf 118}, no 563, (1995).
%
\bibitem{gw93} F.~Gesztesy and R.~Weikard, \textit{Spectral 
deformations 
and soliton equations}, in
{\it Differential Equations with Applications to 
Mathematical Physics},
W.~F.~Ames, E.~M.~Harrell II, J.~V.~Herod (eds.),
Academic Press, Boston, 1993, pp.~101--139.
%
\bibitem{gw96} 
F.~Gesztesy and R.~Weikard, \textit{Picard potentials and Hill's
equation on
a torus}, Acta Math. {\bf 176}, 73--107 (1996).
%
\bibitem{gw98} 
F.~Gesztesy and R.~Weikard, \textit{Elliptic algebro-geometric 
solutions 
of the KdV and AKNS hierarchies -- an analytic approach}, 
Bull. Amer. 
Math. Soc. {\bf 35}, 271--317 (1998).
%
\bibitem{GW98} 
F.~Gesztesy and R.~Weikard, \textit{A characterization of 
all elliptic
algebro-geometric solutions of the AKNS hierarchy}, Acta 
Math. {\bf 181}, 
63--108(1998).
%
\bibitem{j46} 
C.~G.~T.~Jacobi, {\it \" Uber eine neue
Methode zur Integration der
hyperelliptischen Differentialgleichungen
und \" uber die rationale Form
ihrer vollst\" andigen algebraischen
Integralgleichungen\/}, J.~Reine
Angew.\ Math.\ {\bf 32}, 220--226 (1846).
%
\bi{kr92}
B.~G.~Konopelchenko and C.~Rogers, {\it B\"acklund and 
reciprocal
transformations: Gauge connections}, in {\it Nonlinear 
Equations in the
Applied  Sciences}, W.~F.~Ames and C.~Rogers (eds.), Academic 
Press, Boston, 1992, pp.~317--362.
%
\bi{MS91} V.~B.~Matveev and M.~A.~Salle, {\it Darboux 
Transformations and Solitons}, Springer, Berlin, 1991.
%
\bibitem{Mc79} H.~P.~McKean, {\it Theta functions, solitons, 
and singular curves}, in {\it Partial Differential Equations 
and Geometry}, C.~I.~Byrnes (ed.), Marcel Dekker, New York, 
1979, pp.~237--254. 
%
\bibitem{mc85} H.~P.~McKean, {\it Variation on a theme of
Jacobi\/}, Commun.\ Pure
Appl.\ Math.\ {\bf 38}, 669--678 (1985).
%
\bibitem{mc86} H.~P.~McKean, 
{\it Geometry of KdV(1): Addition and the
unimodular spectral classes},
Rev. Mat. Iberoamericana {\bf 2}, 235--261 (1986).
%
\bibitem{mc87} H.~P.~McKean, 
{\it Geometry of KdV(2): Three examples},
J. Statist. Phys. {\bf 46}, 1115--1143 (1987).
%
\bibitem{mw97}
D.~McRae and R.~Weikard, {\it Theta functions on singular 
hyperelliptic surfaces}, preprint, 1997.
%
\bi{MEKL95} P.~D.~Miller, N.~M.~Ercolani, I.~M.~Krichever, 
and C.~D.~Levermore, {\it Finite genus solutions to the 
Ablowitz-Ladik equations}, Commun. Pure Appl. Math. {\bf 48}, 
1369--1440 (1995).
%
\bibitem{Mi68} R.~M.~Miura, {\it Korteweg-de Vries equation 
and generalizations.~I. A remarkable explicit nonlinear
transformation}, J. Math. Phys. {\bf 9}, 1202--1204 (1968).
%
\bibitem{ohm88}
M.~Ohmiya, {\it On the Darboux transformation of the second
order differential operator of Fuchsian type on the 
Riemann sphere},
Osaka J. Math., {\bf 25}, 607--632 (1988).
%
\bibitem{ohm95}
M.~Ohmiya, {\it KdV polynomials and $\Lambda$-operator},
Osaka J. Math., {\bf 32},  409--430 (1995).
%
\bibitem{ohm97}
M.~Ohmiya, {\it Spectrum of Darboux transformation of 
differential operator}, preprint,  Tokushima University, 1997.
%
\bibitem{om93}
M.~Ohmiya and Y.~P.~Mishev,
{\it Darboux transformation and $\Lambda$-operator},
J. Math. Tokushima Univ., {\bf 27}, 1--15 (1993).
%
\bibitem{Pr96} E.~Previato, {\em Seventy years of spectral 
curves:
1923--1993}, in {\it Integrable Systems and Quantum Groups\/},
M.~Francaviglia and S.~Greco (eds.), Lecture Notes in Math.,
Vol.~1620, Springer, Berlin, 1996, pp.~419--481.
%
\bibitem{Pr98} E.~Previato, {\em Burchnall-Chaundy bundles}, 
in {\it Algebraic Geometry}, P.~E.~Newstead (ed.), Marcel Dekker, 
New York, 1998, pp.~377--383. 
%
\bibitem{sc78}
U.-W.~Schmincke,
{\it On Schr\"odinger's factorization method
for Sturm--Liouville operators},
Proc. Roy. Soc. Edinburgh Sect. A, {\bf 80}, 67--84
(1978).
%
\bibitem{SW85} G. Segal and G. Wilson,
{\it Loop groups and equations of KdV
type}, Publ. Math. IHES {\bf 61}, 5--65 (1985).
%
\bi{VS93} A.~P.~Veselov and A.~B.~Shabat, {\it Dressing chains 
and the spectral theory of the Schr\"odinger operator}, Funct. Appl.
{\bf 27}, 81--96 (1993). 
%
\bibitem{Wi85} G.~Wilson, {\em Algebraic curves and soliton
equations},  in {\it Geometry Today\/},  E.~Arbarello, 
C.~Procesi,
and E.~Strickland (eds.), Birkh\"auser, Boston, 1985, 
pp.~303--329.
\end{thebibliography}
\end{document}